\documentclass{aa}  
\usepackage{amsmath}
\usepackage{natbib}
\usepackage{units}
\usepackage{color}

\usepackage{graphicx}
\usepackage{txfonts}

\newcommand{\beq}{\begin{equation}}
\newcommand{\eeq}{\end{equation}}
\newcommand{\AU}{\text{ au}}

\begin{document}
	\title{Particle dynamics in discs with turbulence generated by the vertical shear instability}
	\author{Moritz H. R. Stoll
	\and
        Wilhelm Kley
	}

	\institute{
	Institut f\"ur Astronomie und Astrophysik, Universit\"at 
	T\"ubingen, Auf der Morgenstelle 10, D-72076 T\"ubingen, Germany\\
	\email{moritz.stoll@student.uni-tuebingen.de,wilhelm.kley@uni-tuebingen.de}\\
	}

 	\date{}

\abstract
{Among the candidates for generating turbulence in accretion discs in situations with low intrinsic ionization the
vertical shear instability (VSI) has become an interesting candidate, as it relies purely on a vertical gradient in the angular
velocity. Existing numerical simulations have shown that $\alpha$-values a few times $10^{-4}$ can be generated.}
{The particle growth in the early planet formation phase is determined by the dynamics of embedded dust particles.
Here, we address in particular the efficiency of VSI-turbulence in concentrating particles in order to generate overdensities and low
collision velocities.}
{We perform three-dimensional (3D) numerical hydrodynamical simulations of accretion discs around young stars that include radiative transport and
irradiation from the central star. The motion of embedded particles within a size range of a fraction of mm up to several m is followed using
standard drag formula.}
{We confirm that under realistic conditions the VSI is able to generate turbulence in full 3D protoplanetary discs. 
The irradiated disc shows turbulence within 10 to 60 \AU. The mean radial motion of the gas is such that it is directed inward near the
midplane and outward in the surface layers.
 We find that large particles drift inward with the expected speed, while small
particles can experience phases of outward drift. 
Additionally, the particles show bunching behaviour with overdensities reaching 5 times the average value,
which is strongest for dimensionless stopping times around unity.
}
{Particles in a VSI-turbulent discs are concentrated in large scale turbulent eddies and show low relative speeds that allow for growing collisions.
The reached overdensities will also allow for the onset streaming instabilities further enhancing particle growth. The outward drift for small
particles at higher disk elevations allows for the transport of processed high temperature material in the Solar System to larger distances.}
   \maketitle
%

\section{Introduction}
To drive mass flow in accretion discs an anomalous source of angular momentum is required \citep{2002apa..book.....F}. 
A strong candidate is the magneto-rotational instability (MRI), which gives rise to turbulent magnetohydrodynamical (MHD) flows
that create an outward angular momentum transport discs \citep{1998RvMP...70....1B}.
Driven by magnetic fields, the MRI requires a sufficient level of ionization to sustain a turbulent state within the disc.
However, protoplanetary discs have only a very low temperature regime and insufficient thermal ionization.
Even considering external sources of ionization there appears to be a region of insufficient ionization level such that the
MRI cannot operate, as shown by resistive MHD simulations including radiative transport \citep{2012MNRAS.420.2419F}.
Hence, there may exist a dead zone somewhere between $2-20\AU$ \citep{Armitage2011ARA&A..49..195A},
where the MRI can only produce very weak turbulence.  Recent simulations which included, in addition to Ohmic resistivity,
also ambipolar diffusion have even shown no signs of turbulence at all in this region \citep{2015ApJ...801...84G}. The Hall effect creates strong winds in the surface of the disc and may even reintroduce angular momentum transport in the dead zone, but this depends on the sign of the magnetic field \citep{Bai2014ApJ...791..137B,Bai2015ApJ...798...84B}.
Thus another origin of instability inside the dead zones is warranted to drive accretion in protoplanetary discs.

As an alternative to the MRI, different examples of purely hydrodynamic instabilities in discs have been suggested, such as 
the  gravitational instability \citep{1987MNRAS.225..607L}, the convective instability \citep{1988ApJ...329..739R},
or the baroclinic instability \citep{2003ApJ...582..869K}, but they do not operate under general conditions. 
One possibility, that has attracted recently more attention is the vertical shear instability (VSI) suggested for accretion discs by
\citet{Urpin2003A&A...404..397U}. The mechanism was first examined in relation to differential rotating stars 
\citep{Goldreich1967,1968ZA.....68..317F}, and it is also known as Goldreich-Schubert-Fricke instability.
In the context of discs first simulations have been carried out by \citet{2004A&A...426..755A}.
While there is a much larger radial gradient in the angular velocity, $\Omega$, to feed instabilities, most instabilities cannot 
overcome the stabilising effect of rotation. In the context of the VSI, it is the vertical shear in $\Omega$ created by a
radial temperature gradient that allows the disc to become unstable. The numerical work of \citet{Nelson2013MNRAS.435.2610N} showed that 
a small turbulent $\alpha$-value in the range of a few $10^{-4}$ was possible for isothermal discs.
For non-isothermal discs \citet{Nelson2013MNRAS.435.2610N} point out that radiative cooling (diffusion) and viscosity will reduce
the instability, and they developed a theoretical model describing the initial vertical elongated modes destabilising the disc.
In simulations that included full radiative transport \citet{Stoll2014A&A...572A..77S} showed that for situations typical
in protoplanetary discs a sustained VSI was possible providing an $\alpha \sim 10^{-4}$. They found the development of
a global wave pattern within the disc whose wavelength was determined partly by viscous effects.
Later, \citet{Barker2015MNRAS.450...21B} analyzed the VSI through linear analyses of locally isothermal discs
and support the modal behaviour seen in the non-linear simulations by \citet{Nelson2013MNRAS.435.2610N} and \citet{Stoll2014A&A...572A..77S}.
They also stress the importance of viscosity to set the smallest length scale. 

Recently, \citet{Lin2015ApJ...811...17L} shed light on the cooling requirements of the VSI, that arise because the
VSI has to compete with the stabilising vertical buoyancy. Their theoretical models predict activity in regions with
large cooling time only for very large wavenumbers. Thus the VSI is limited by the cooling time on large scales and
by viscosity on small scales. With this in mind they predict VSI activity for typical disc models only between 5 and 100 au.

We test this idea by expanding our previous work \citep{Stoll2014A&A...572A..77S} where we used a self-consistent 
radiation transport module and a vertically irradiated disc in the simulations. 
Here, we treat the irradiation in a more realistic way as originating from the central star, and show that even under this condition
the VSI can be sustained.

The turbulence in protoplanetary discs is also critical for the initial dust evolution that leads eventually to planet formation.
Numerical simulations performed by \citet{Johansen2005a} and \citet{Fromang2005MNRAS.364L..81F} showed that particles
can be caught in the local pressure maxima generated by the MRI turbulence. This clustering of dust particles can then 
trigger a streaming instability \citep{2005ApJ...620..459Y} that will lead to further clustering and subsequently 
to the formation of $km$-planetesimals. 
In the context of the VSI the large scale velocity patterns of the corrugation mode
promises interesting behaviour for embedded dust grains and larger particles.
To investigate the impact of the VSI modes on the dust particles we add particles into our disc model and follow their
dynamical evolution. 

The paper is organized as follows. In Sect.~\ref{sect:setup} we present our numerical and physical setup.
We present a detailed analysis of an isothermal disc model in Sect.~\ref{sect:isodisc}, and we discuss in Sect.~\ref{isothermal} 
the results for the particle evolution is this model. In Sect.~\ref{sec:viscosity} we describe the results of a viscous model.
The simulations with radiation transport and stellar irradiation are presented in Sect.~\ref{radiative} and in Sect.~\ref{conclude} we conclude.
 
\section{The model setup}\label{sect:setup}
We use the same equations and physical disc setup, as described in detail in our first paper \citep{Stoll2014A&A...572A..77S}
and give here only a very brief outline.
In summary, for the integration of the hydrodynamical equations we use \textit{PLUTO} a publicly available
code, based on a Godunov scheme for viscous hydrodynamical flow \citep{2007ApJS..170..228M}, extended by flux-limited diffusion 
module for radiation transport and a ray-tracing method for stellar irradiation \citep{2013A&A...559A..80K}. 
Having used a two-dimensional, axisymmetric disc setup in our previous work, we now extend the computational domain in the azimuthal direction to three dimensions (3D) and add particles to the flow.
For this purpose we added a particle solver based on the method by \citet{Bai2010ApJS..190..297B},
that we describe in the next section.

\subsection{Particle Solver}
We use Lagrangian particles with drag and gravitation:
\begin{equation}
\label{eq:particles}
    \frac{d\vec{v}_p}{dt} = \vec{a} =  \vec{f} + \frac{\vec{v}_p -\vec{u} }{t_s} \,,
\end{equation}
where $\vec{f}$ is an acceleration due to an external force, here the gravitation of the star. $\vec{v}_p$ and $\vec{u}$ are the particle and gas velocity and $t_s$ is the stopping time. 

We treat all particles as if they were in the Epstein regime where the mean free path of the gas molecules
is typically larger than the particle cross section \citep{1924PhRv...23..710E}. The stopping time is then
\begin{equation}
\label{eq:stop}
    t_s = \frac{r_p \rho_p }{\rho_g \sqrt{8/\pi}c_s} \,,
\end{equation}
where $r_p$ is particle radius, $\rho_p$ the particle bulk density, $\rho_g$ the gas density, and $c_s$ is the sound speed.
In addition we will use $\tau_s = t_s \Omega_K$ for the dimensionless stopping time. 

To solve the equation of motion of the particles (\ref{eq:particles}) we follow \citet{Bai2010ApJS..190..297B}.
If the stopping time is larger than the time step, $\Delta t$, of the simulation, we solve the semi-implicit equations
\begin{eqnarray*}
    \vec{x'} &=& \vec{x}^{(n)} + \frac{\Delta t}{2} \vec{v}_p^{(n)} \,, \\
    \vec{v}_p^{(n+1)} &=& \vec{v}_p^{(n)} + \Delta t \vec{a}\left[ ( \vec{v}_p^{(n)} + \vec{v}_p^{(n+1)})/2,\vec{x'}\right]  \,, \\
    \vec{x}^{(n+1)} &=& \vec{x'} + \frac{\Delta t}{2} \vec{v}_p^{(n+1)} \,,
\end{eqnarray*}
where $n$ denotes the timestep level, and $\vec{x'}$ an intermediate position of the particle. The 
particle acceleration, $\vec{a}$, is a function of the particle velocities, $\vec{v}_p$, see Eq.~(\ref{eq:particles}).

For stopping times smaller than the time step we solve the following implicit equation, where the velocity update does not depend on the old velocity. 
\begin{eqnarray*}
    \vec{x'} &=& \vec{x}^{(n)} + \Delta t \vec{v}_p^{(n)} \\
    \vec{v}_p^{(n+1)} &=& \vec{v}_p^{(n)} + \frac{\Delta t}{2} (\vec{a} [ \vec{v}_p^{(n+1)},\vec{x'}] 
    +  \vec{a} [ \vec{v}_p^{(n+1)}-\Delta t\vec{a} ( \vec{v}_p^{(n+1)},\vec{x'}),\vec{x}^{(n)}]) \\
    \vec{x}^{(n+1)} &=& \vec{x}^{(n)} + \frac{\Delta t}{2} (\vec{v}_p^{(n)} + \vec{v}_p^{(n+1)})
\end{eqnarray*}
This allows the drag force to damp the particle velocity without unphysical oscillations.
The test simulations to verify the correct implementation of the particle solver are described
in the appendix in Sect.~\ref{appsect:particles}.

\subsection{Physical setup}\label{physical}
In order to study the importance of radiative effects we decided to perform first a sequence of isothermal simulations.
We describe briefly the setup of our fiducial disc model, which consists of a 3D isothermal model that simulates one eighth
(covering $45^\circ$ in azimuth)
of the disc at a resolution of $1024\times256\times64$ (in $r, \theta, \phi$) without viscosity. 
The full radiative model will be described in section \ref{radiative}.

We use the same general disc setup as in \citet{Stoll2014A&A...572A..77S}, where we started with a disc in force equilibrium,
\begin{equation}
  \rho(R,Z) = \rho_0 \left( \frac{R}{R_0} \right)^p \,  \exp{ \left[ \frac{G M}{c_s^2}\left( \frac{1}{\sqrt{R^2 + Z^2}} - \frac{1}{R} \right) \right]} \,,
\label{eq:rho2d}
\end{equation}
with a temperature that is constant on cylinders
\begin{equation}
  T(R,Z) = T_0 \left( \frac{R}{R_0} \right)^q \,.
\label{eq:Temp0}
\end{equation}
Here, $c_s = \sqrt{p/\rho}$ denotes the isothermal sound speed,  and
$H = R h = c_s / \Omega_K$ is the local pressure scale height of the accretion disc. 
We use (R,Z,$\phi$) for cylindrical coordinates and ($r$,$\theta$,$\phi$) for the spherical coordinates that we use in our simulations. 
Typical values for the density and temperature exponents are $p = -1.5$ and $q = -1$.

The computational domain of the fiducial model is limited to $r = 2 - 10 \AU$ in the radial and $\pm 5$ scale heights in the meridional direction.
In contrast to  \citet{Stoll2014A&A...572A..77S}, where we only simulated 2D, axisymmetric discs, we use here in the azimuthal direction 
one eighth of the full disc ($\phi_{max} = \pi/4$) to capture the complete 3D physics of the turbulent disc.

\begin{figure}[bt]
\begin{center}
\includegraphics{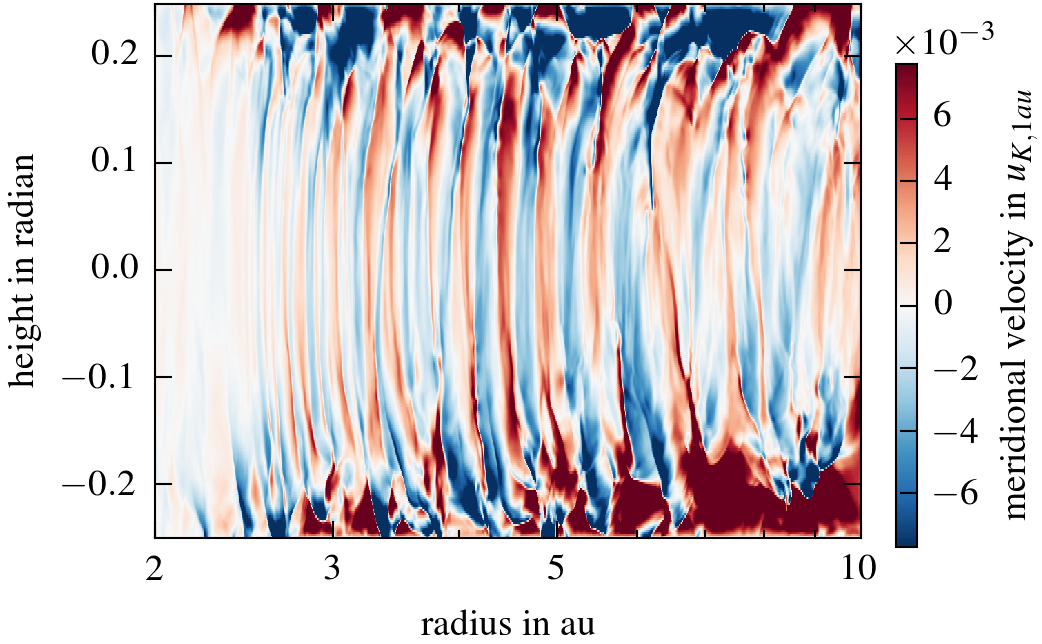}
\end{center}
\caption{The velocity in the meridional direction, $u_\theta$, in units of the Kepler velocity for the fiducial isothermal run 
 without viscosity and a resolution of $1024\times256\times64$. Displayed is the quasi-stationary state after 1200 years.} 
\label{fig:vx2_rtheta}
\end{figure}

\section{The turbulent isothermal disc}
\label{sect:isodisc}
Before embedding the particles we first describe the turbulent properties of the VSI for several disc models.
We start with the fiducial isothermal and inviscid model for one eighth of the disc and compare this below 
to models with viscosity and a larger azimuthal domain. 
In the subsequent section we use the fiducial model for the simulations with particles. 
Finally, in the last section we will also show the results of a radiative, irradiated disc model. 

As shown in previous numerical simulations of the VSI the turbulent state is characterized by vertically elongated
flow structures as shown in Fig.~\ref{fig:vx2_rtheta} for the meridional velocity, for more details see
\citet{Nelson2013} and \citet{Stoll2014A&A...572A..77S}.
To characterize the turbulence of the disc we measure the Reynolds stress that we define here  as 
\begin{equation}
    T_{r,\phi} = \left< \rho u_r \delta u_{\phi} \right>_{\phi,t} \,, 
    \label{eq:Reynoldsstress}
\end{equation}
where $\delta u_{\phi}(r,\theta,\phi) = u_{\phi}(r,\theta,\phi) - \left< u_{\phi} (r,\theta,\phi)\right>_{t}$ 
is the deviation of the angular velocity from the (time averaged) mean azimuthal velocity. We denote averages taken over
certain variables by $\left< \, \right>$ with the appropriate indices.
In this definition, $T_{r,\phi}$ is a 2D array in $r, \theta$.  
The time average, $ \left< u_{\phi} \right>_t$, needed in Eq.~(\ref{eq:Reynoldsstress}) is in general not known in advance but one can rewrite
the Reynolds stress as
\begin{equation}
    T_{r,\phi} 
   = \left< \rho u_ru_{\phi} \right>_{\phi,t} - \left< \rho  u_{r} \right>_{\phi,t} \left< u_{\phi} \right>_{\phi,t} \,,
    \label{eq:trp1}
\end{equation}
where the right hand side can in principle be calculated ''on the fly'' during the simulation.  
For convenience, we chose to store $\phi$-averaged 2D data sets at regular time intervals, and then average these over time 
to obtain $T_{r,\phi}(r, \theta)$.

From these we calculate the dimensionless $\alpha$-parameter as a function of radius 
\begin{equation}
    \alpha(r) = \frac{\left<T_{r,\phi} \right>_{\theta}}{\left< P \right>_{\theta}},
    \label{eq:reynolds}
\end{equation}
where $P=\left< p \right>_{\phi,t}$ is the azimuthal and time averaged pressure.
For the vertically dependent $\alpha(z)$-parameter at a certain radius $r_c$,
\begin{equation}
    \alpha_{r_c}(z) = \frac{\left<T_{r,\phi} \right>_{r}}{\left< P \right>_{r}},
    \label{eq:reynoldsvertical}
\end{equation}
we integrate only over a small radial domain around the desired radius, $r_c$.
This averaging procedure to calculate $\alpha$ has to be used for general discs, for example the radiative discs below.

\begin{figure}[tb]
    \centering
    \includegraphics{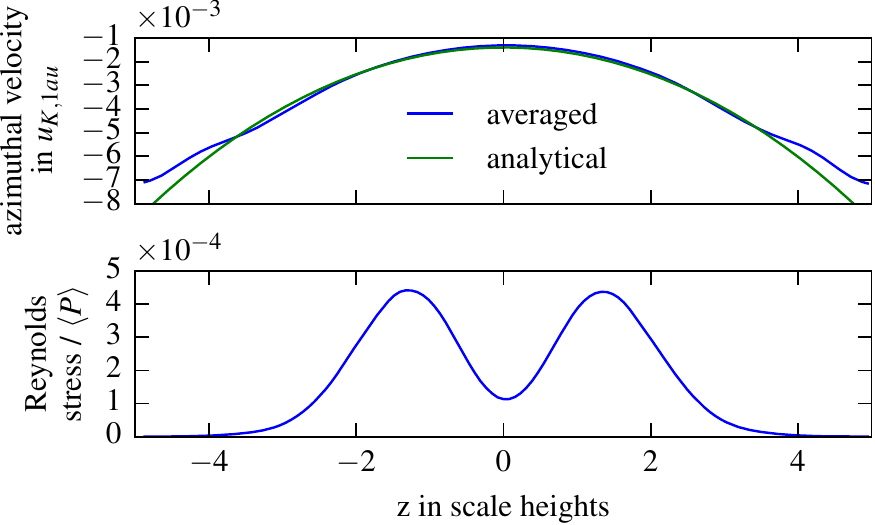}
    \caption{The averaged azimuthal velocity compared to the analytical velocity  (upper panel) and $\alpha(z)$ (lower panel).
    For the green (analytical) curve Eq.~(\ref{eq:omega}) has been used for the mean value of $u_\phi$. 
    The blue curve shows the average from 1000 to 1800 years over 800 time levels and radially from $4.5$au to $5.5$au.}
    \label{fig:velocity}
\end{figure}

However, for the isothermal simulations, one can approximate the time averaged $\left< u_{\phi} \right>_t$ in the calculation of the Reynolds stress, by
the analytically calculated solution for the equilibrium angular velocity \citep{Nelson2013}, that can be obtained from the initial equilibrium disc setup 
\begin{equation}
\Omega(R,Z) = \Omega_K \left[ (p +q)\left( \frac{H}{R} \right)^2 + (1+q) - \frac{qR}{\sqrt{R^2 + Z^2}} \right]^{\frac{1}{2}} \,,
\label{eq:omega}
\end{equation}
where $\Omega_K = \sqrt{G M_{\odot}/R^3}$ is the Keplerian angular velocity. In Fig. \ref{fig:velocity} (upper panel) we can see that this is indeed a valid approximation. Aside from reducing noise in $\alpha$, since it no longer depends on the time averaged velocity, this has the further advantage of allowing us to directly calculate the $\alpha$-parameter at each timestep and we thus only need to store 1D arrays, which we then can later average over arbitrary timespans.

\begin{figure}[bt]
\begin{center}
\includegraphics{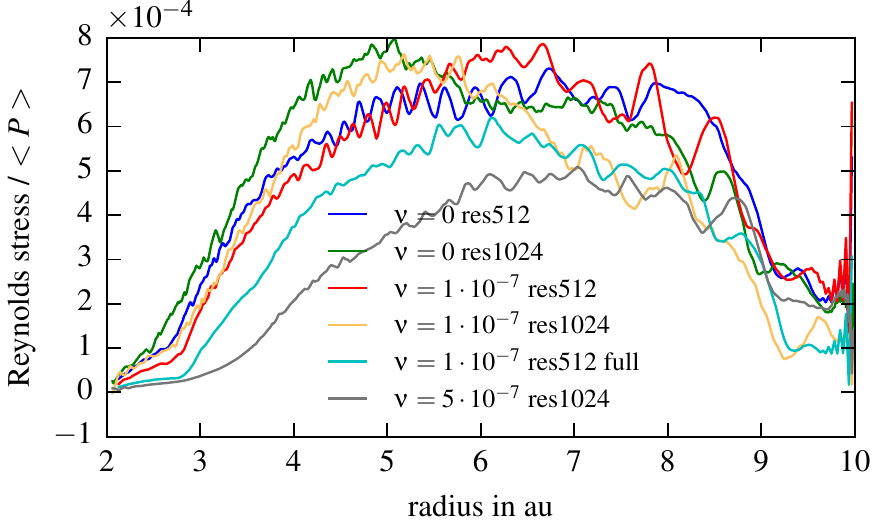}
\end{center}
\caption{Comparison of the radial $\alpha(r)$ obtained for different isothermal simulations used in this paper. 
Values are averaged from 1000 to 1800 years. All simulations except 'full' are for one eighth of a complete disc.
The specified resolution refers to the number of radial grid cells in the simulations, where 'res1024' denotes the resolution of our fiducial model and 'res512' has half the resolution in all spatial directions. In addition to the inviscid fiducial model we ran also simulation with non-zero viscosity, and the labels refer to the dimensionless value of the constant kinematic viscosity coefficient $\nu$.
}
\label{fig:alpha_iso}
\end{figure}

\begin{figure}[bt]
\begin{center}
\includegraphics{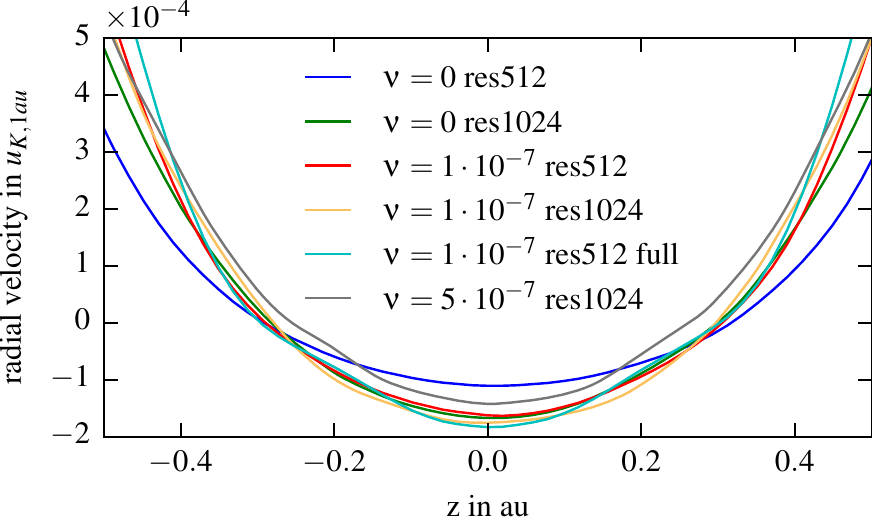}
\end{center}
\caption{The radial velocity $u_r$ (in units of the Kepler velocity at $1\AU$) for the different isothermal simulations evaluated at $5\AU$. 
 Shown are the same models as in the previous Fig.~\ref{fig:alpha_iso}, averaged again from 1000 to 1800 years.
Negative velocities correpond to inflow towards the star.}
\label{fig:vr_iso}
\end{figure}

In Fig. \ref{fig:alpha_iso} we compare the radial $\alpha$-parameter for different isothermal simulations. 
The averaging was done from 1000 years to 1800 years with 800 snapshots taken at regular intervals. Shown are cases with different resolutions, where the labels indicate the radial number of grid cells, and also viscous models.  The dimensionless viscosity is given in units of $(u_{K,1\AU} \cdot 1 \AU) $, where $u_{K,1\AU}$ is the Kepler velocity at 1\AU.
We can see that in contrast to the 2D simulations in \citet{Stoll2014A&A...572A..77S}, where the $\alpha$-parameter differed by more than 50\% when we doubled the resolution, these 3D simulations show no clear dependence on resolution, with all curves being within 10\% of each other at $6\AU$.
We ran a further, shorter simulation without viscosity and double resolution of $2048 \times 512 \times 128$ and could not see a difference in wavelength or Reynolds stress in the early equilibrium phase from 500 to 800 years. Additionally we see only a weak dependence on viscosity with noticeable differences beginning with the largest kinematic viscosity, which starts to suppress the VSI in the inner region. A further increase in viscosity would suppress the instability completely, as shown already in \citet{Nelson2013MNRAS.435.2610N}.

Since we are interested in the particle drift we also show in Fig. \ref{fig:vr_iso} the radial velocity in the disc at $5\AU$ as a function of height. We averaged from 1000 years to 1800 years, which is roughly the quasi-stationary phase. The radial velocity is inwards in the midplane and outwards in the corona. There are only minor differences in height were the direction of the flow changes. Interestingly, this profile is opposite to that of a laminar viscous flow, where it is outwards in the midplane, and inward near the disc surfaces \citep{1984SvA....28...50U,1992ApJ...397..600K}.
Our findings are in agreement with results of isothermal MHD simulations of global turbulent accretion discs without a net magnetic vertical flux
that also show gas inflow near the disc midplane and outflow in the disc's surface layers \citep{2011ApJ...735..122F}.
Within the framework of viscous discs the vertical variation of $\alpha$ (shown in the lower panel of Fig.~\ref{fig:velocity})
will play a role in determining the $u_r(z)$-profile, see also \citet{1992ApJ...397..600K} and \citet{Takeuchi2002ApJ...581.1344T}, who studied the 
the $u_r(z)$-profile for constant $\alpha$.

\subsection{3D-simulation: full disc}
In this subsection we present a full disc, meaning an azimuthal domain from 0 to $2\pi$ in contrast to the 0 to $\pi/4$ of our fiducial model, 
in order to check the validity of our results obtained from calculations with the reduce domain.
Since this full simulation is computationally expensive, we used a lower resolution of $512\times128\times512$.
The full simulation is compared to the one eighth simulation of the disc with a resolution of $512\times128\times64$,
which has the same azimuthal extent as the fiducial model, but not the resolution.  We added a small dimensionless physical viscosity, $\nu = 10^{-7}$, to be independent of the unknown numerical viscosity, allowing better comparison with other simulations.   
\begin{figure}[t]
    \centering
    \includegraphics{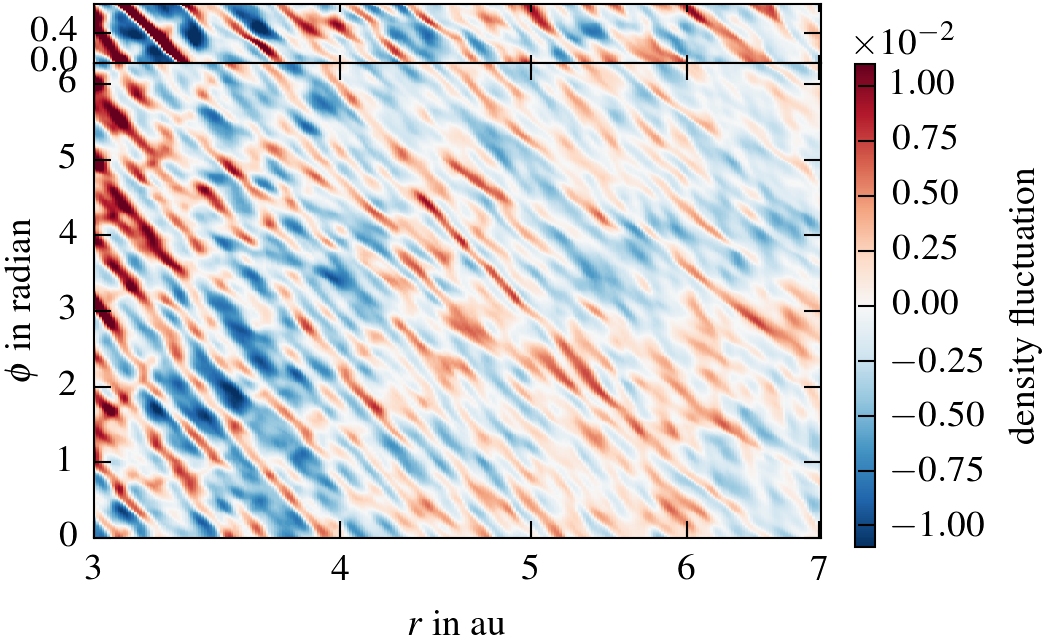}
    \includegraphics{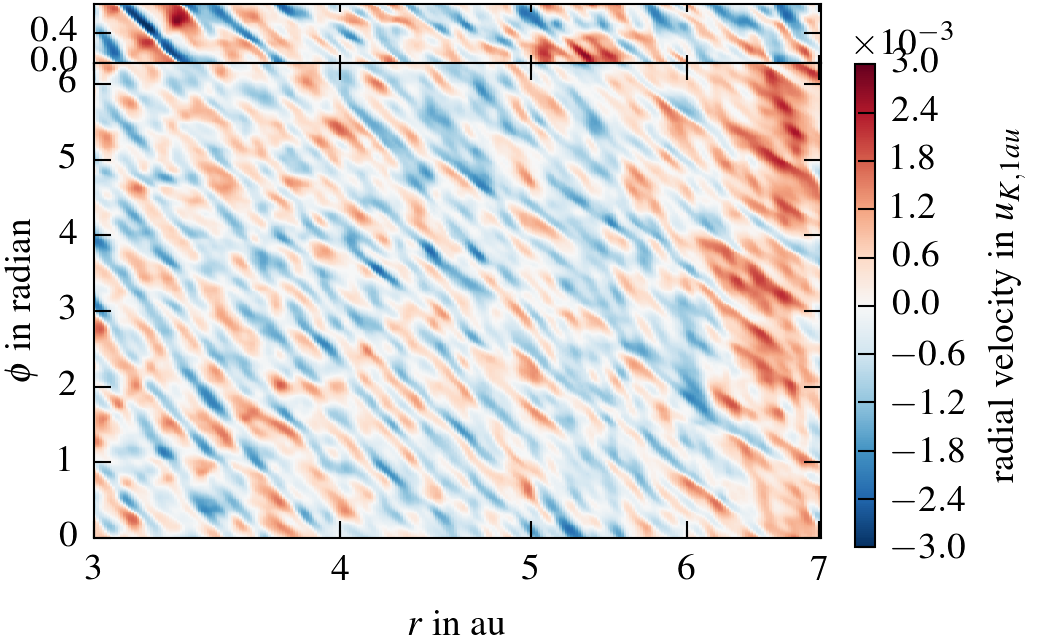}
    \includegraphics{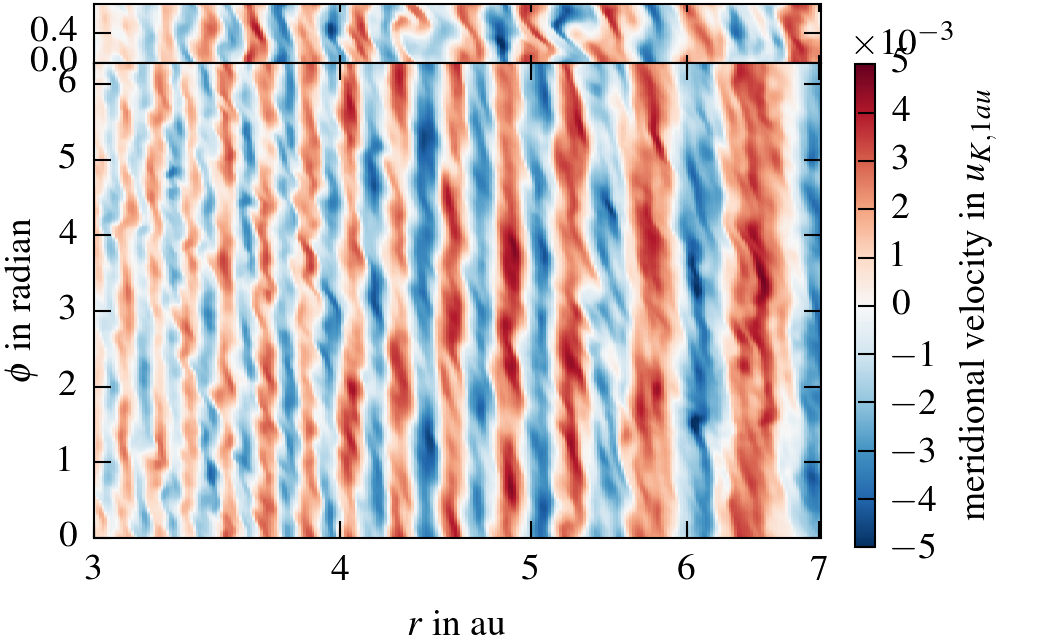}

    \caption{The fluctuations of density, the radial and vertical velocity (from top to bottom) 
     in the midplane of the disc with 3 to $7\AU$ after 1700 years.
     The top part in each panel refers to the fiducial model with one eighth of the full azimuthal domain while the full model
     is shown in the lower part of each panel.}
    \label{fig:corr2plots}
\end{figure}
To give an impression of possible differences in the flow structures between the full and fiducial
model we display in Fig.~\ref{fig:corr2plots}, from top to bottom, the fluctuations of the density, and the radial and vertical velocity components in the midplane of the disc, where the top inset in each panel refers to the fiducial model with $\phi_{max}=\pi/4$ and the bottom part to the full disc.   The two top panels for the density and radial velocity clearly show non-axisymmetric, wave-like features in the disc. The bottom panel seems to indicate a more axisymmetric structure which is a result of the VSI eigenmode dominating the vertical motion in the disc, as seen above in Fig.~\ref{fig:vx2_rtheta}.

\begin{figure}[t]
    \centering
    \includegraphics{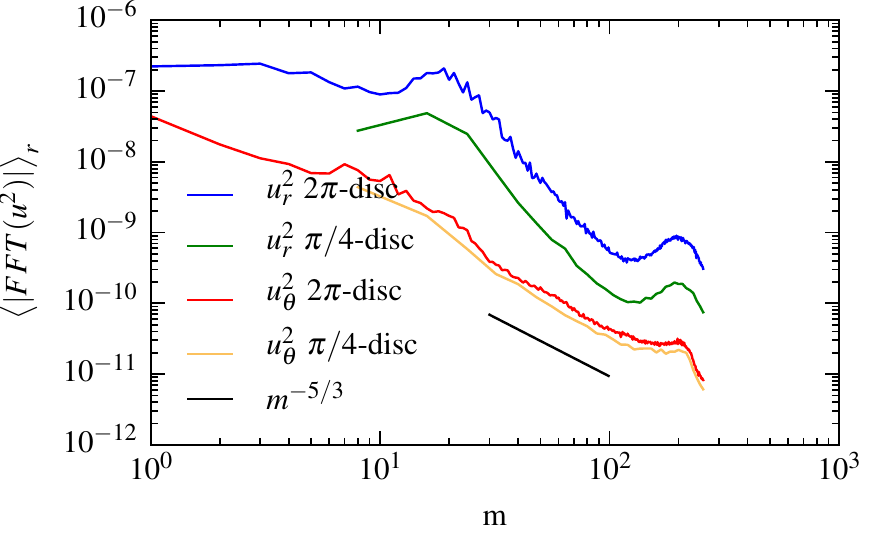}
    \includegraphics{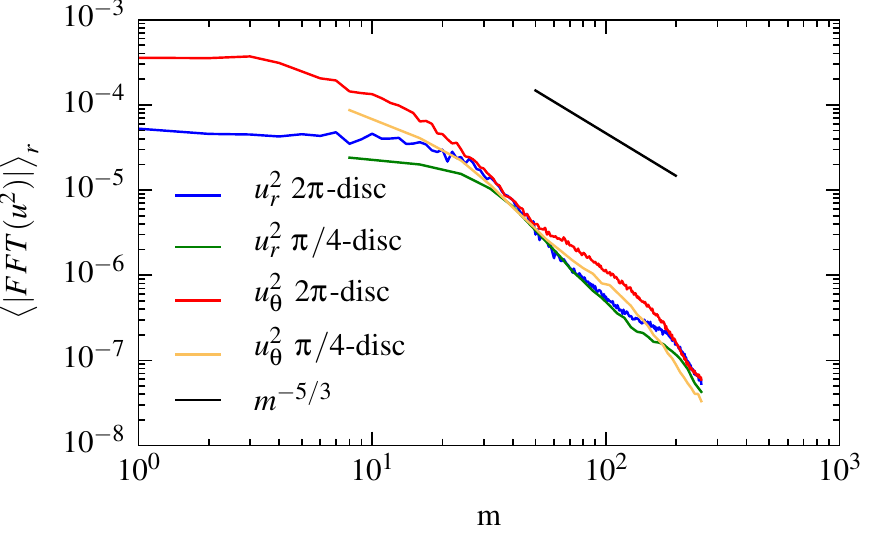}
    \caption{Power spectrum for the different kinetic energy components along the azimuthal direction in the disc midplane averaged
    over the radial direction. The black line shows the Kolmogorov spectrum decaying with $|u(m)|^2 \propto m^{-5/3}$
    where $m$ denotes the azimuthal wave number.
    The top panel refers to the growth phase after 200 years and the lower panel to the quasi-stationary phase after 1700 years.}
    \label{fig:fourier_step17}
\end{figure}

To analyze the turbulent structure in more detail we calculated azimuthal power spectra of the radial and vertical kinetic energy fluctuations in the disc midplane for two different times, spatially averaged from 3 to $7\AU$.
We can see in Fig.~\ref{fig:fourier_step17} that during the initial growth phase (top panel) the instability is driven at two length scales.
The smaller one, at azimuthal wave numbers $m \approx 200$, is most likely due to the initial noise as given by the finite discretization which is enhanced by the growth phase of the instability. The larger one (at $m \approx 10$) is on the scale of the wavelength of the strongest growing VSI mode. This feature is also visible in the azimuthal direction even though the VSI should be axisymmetric. We speculate that this is due to a disturbance created in the radial direction by a Kelvin-Helmholtz instability that is sheared into the azimuthal direction. 
We can also see that in the quasi stationary phase (bottom panel) the turbulence decays faster than Kolmogorov turbulence for wavenumbers around $m \approx 20$, which is the scale at which the VSI is driven. There the energy is concentrated in the VSI modes and not in the turbulent kinetic energy and thus the model of Kolmogorov decay is not applicable \citep{Dubrulle1992A&A...266..592D}. Only at small scales the vertical kinetic energy fluctuations decay  with a Kolmogorov spectrum. This can be seen in both simulations. From this we infer that the local properties of the turbulence generated by the VSI are very similar for the restricted and full azimuthal domain and is already fully captured by the smaller disc. 

To analyse the locality of the VSI further, we also compare the two point correlation functions for the two disc models.
This is defined as
\begin{equation}
    \xi_f (\Delta r, \Delta \phi) = \left< f(r, \phi ) f(r+\Delta r(r), \phi +\Delta \phi) \right> \,,
    \label{eq:corr2}
\end{equation}
where $f$ is the quantity to be correlated, which has a zero mean, $<f> = 0$.
We evaluate the correlation in the disc midplane and take the radial domain from 3 to $7\AU$, which we treat for this calculation as periodic. The correlations are evaluated on a logarithmic grid, to better capture the properties under investigation.

\begin{figure}[t]
    \centering
    \includegraphics{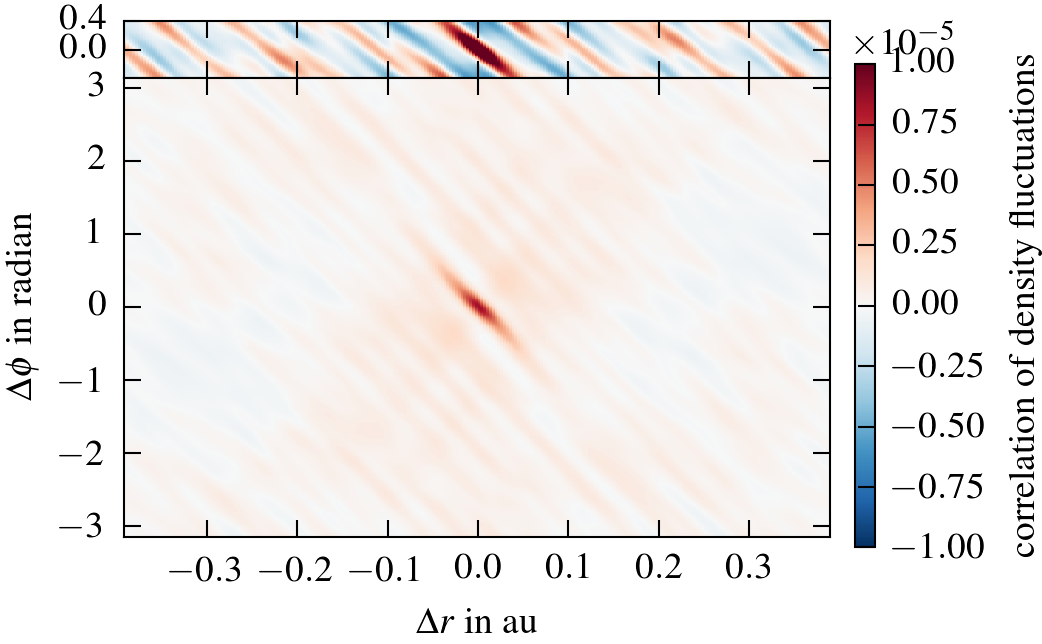}
    \includegraphics{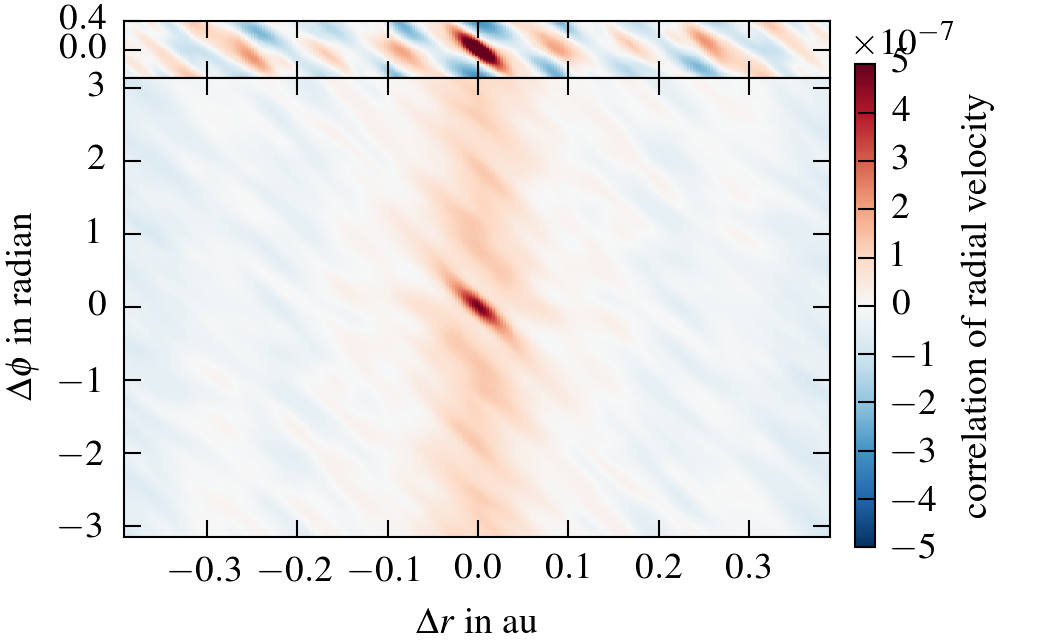}
    \includegraphics{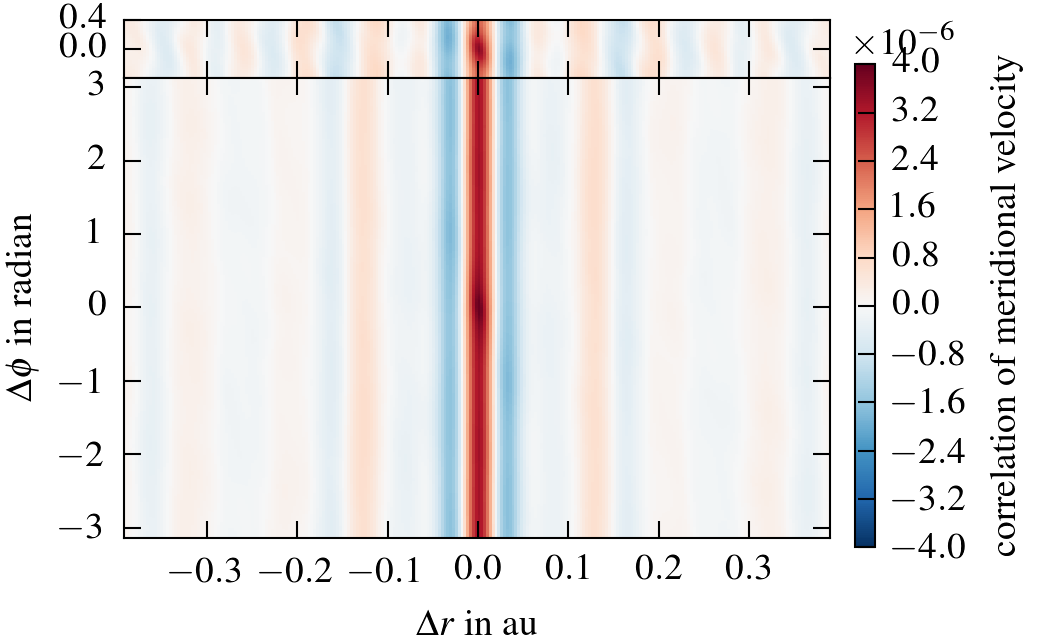}
    \caption{The two point correlation of the density, the radial and vertical velocity (from top to bottom)
     in the midplane of the disc with 3 to $7\AU$ after 1700 years.
     The top part in each panel refers to the fiducial model with one eighth of the full azimuthal domain while the full model
     is shown in the lower part of each panel.}
    \label{fig:corr2_step15vx2}
\end{figure}

In Fig. \ref{fig:corr2_step15vx2} we present the results of the two point correlations for the density fluctuations and the radial and meridional velocity after 1700 years. The fluctuations  are clearly non-isotropic and correlated not only on a local scale but also weakly over the whole domain. We can see again that the smaller domain is a reasonable approximation to the larger domain, even though the correlations are enhanced in the smaller domain, but for the vertical velocity we have a global correlation for both cases. This again strengthens the impression that the fluctuations in azimuthal direction are driven by a Kelvin-Helmholtz instability that feeds off the strongest VSI mode. From this we conclude that our reduced domain captures all of the important physics even though it enforces stronger correlations.\footnote{A simulation with a quarter disc still shows the same enhanced correlations}

The models for different levels of background viscosity  were shown already in Fig. \ref{fig:alpha_iso}. The full model has a slightly smaller $\alpha$, but is apart from this very similar. We also compared the results from the particle motions (not shown), but also found only minor differences. 
Hence, we conclude that we can use the reduced model with $\phi_{max} = \pi/4$ to analyze the motion of embedded particles.

\section{Isothermal discs with dust}\label{isothermal}

The particles are added into the fiducial model after 200 years when the VSI is reaching the quasi-stationary phase at the inner part of the disc. After further 800 years both VSI and particles are then in quasi-equilibrium. The particles are inserted in the midplane, randomly distributed  over the radius and azimuthal angle, with the velocity of the gas at their current position. If a particle leaves the domain in the inner edge we insert it again randomly positioned over $1\AU$ at the outer edge. This ensures that a clump of particles that leaves the domain is sufficiently smoothed out when added again. We add 10,000 particles per size with 20 different sizes, beginning with $1\mu$m up to $3000$m. The Epstein regime is strictly valid only up to sizes of $10$m, and particles larger than that are added as a numerical experiment. Note, that for the isothermal discs the scale for the density is not fixed and thus the particle size regime may be different for different choices of the disc density. This is not the case for the disc with radiation transport as shown in the next section, where the chosen disc mass (the density) fixes the disc temperature, and hence the disc scale height, through specific values of the opacity.

We begin with the results for the radial drift and diffusion of the dust particles, then we discuss the vertical diffusion and finally the relative velocity distribution for colliding particles. We will dicuss our results either in terms of the physical size of the particles or the stopping time.
For the models in this section the correspondence between these can be read off from Table~\ref{tab:difvert}.

\subsection{Radial particle drift}

The radial drift velocity for the dust in the midplane in the Epstein regime is given by \citep{Nakagawa1986Icar...67..375N} 
\begin{equation}
    u_{\rm drift} = \frac{\partial \ln p}{\partial \ln R} \frac{(H/R)^2 u_K}{\tau_s +\tau_s^{-1}} \,.
    \label{eq:radialdrift}
\end{equation}
This is due to the drag force resulting from the difference between the Kepler velocity of the particles and the  gas velocity, which is modified by the pressure support $p$. We use the midplane pressure for our theoretical curves, since most particles are in the midplane, at least the larger ones. Here $\tau_s$ is the dimensionless stopping time (see Eq.~(\ref{eq:stop})) and $u_K$ is the Kepler velocity.

\begin{figure}[bt]
\begin{center}
\includegraphics{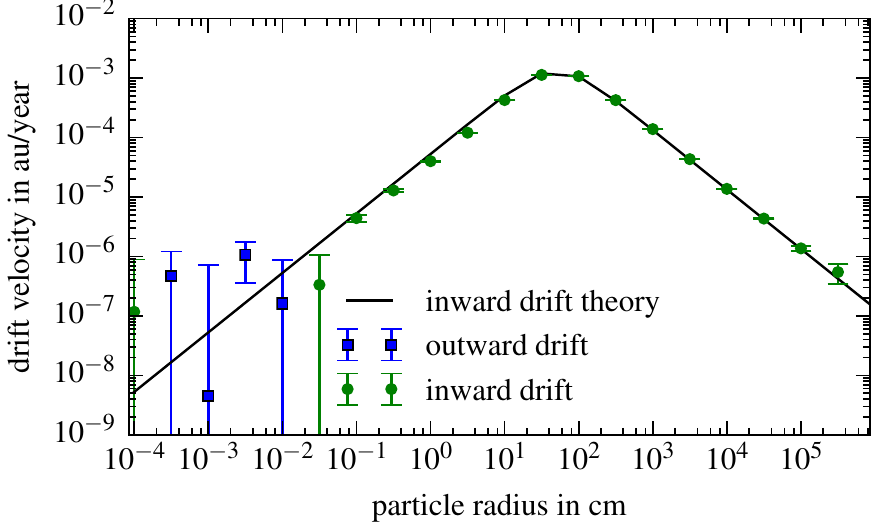}
\end{center}
\caption{The mean radial drift velocity of the dust particles depending on the radius of the particles at $r=5\AU$. 
This is compared to expected drift velocity for pressure supported discs as given by Eq.~(\ref{eq:radialdrift}).
Shown are isothermal disc simulation with resolution $1024\times256\times64$ from 1000 to 2000 years. 
Different colors and symbols are used for inward and outward drift. We estimate the error from the radial diffusion coefficent.}
\label{fig:migration}
\end{figure}

We start by comparing the radial drift of the dust particles in our simulation with this theoretical prediction for the drift velocity in Fig. \ref{fig:migration}. The results from the simulation are extracted by fitting a linear function to the mean radial position of the particles starting with a distance to the star of $5\pm 0.5 \AU$. We fit over the span of 1000 years beginning with 800 years after inserting the particles or over the time the particles need to travel $0.5\AU$, whichever is smaller. The inwardly directed drift is plotted in green (dots), and outward drift in blue (squares). Since the error in the measurement of the velocity stems from the random walk of the particles due to the turbulence, we estimate the error from the radial diffusion coefficent (see Fig. \ref{fig:difRad}).
The error is then given by the half-width of the distribution divided by the square-root of the number of involved particles.

We can see that the speed of large particles is similar to the predictions. 
A difference in speed of approximately 20\% can be seen for the smaller particles in a size range between 0.1 cm to 10 cm. This deviation can be partly attributed to the spread of the particles in the vertical direction, because particles not in the midplane have a larger stopping time (lower gas density), and the prediction is calculated using the stopping time of particles in the midplane. An additional factor is that they can be caught temporarily in small scale vortices \citep{Johansen2005a}.

The drift velocity for the  smallest particles can even be positive for some time intervals. 
This is shown for particles at $5\AU$ in Fig.~\ref{fig:migration}. At distances closer
to the star they clearly drift inwards even though the gas momentum is the same (not shown). We will return to the analysis for those particles later in section \ref{sec:raddif}, where the sign of the migration direction is better constraint.

\begin{figure}[tb]
\begin{center}
\includegraphics{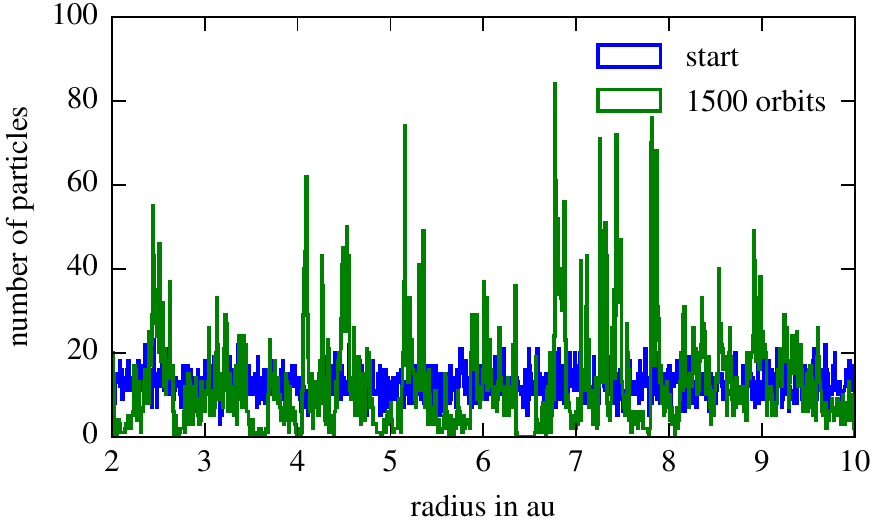}
\end{center}
\caption{Histogram of particles with size of 31cm, after 1500 years (green line) and at the start (blue line, Poisson distributed). We divide the distance from the star from $2$ to $10\AU$ into 800 bins and count the number of particles in each bin. The average number of particles per bin is 12.5.}
\label{fig:grain8}
\end{figure}

An interesting behavior of the particles is shown in the histogram in Fig. \ref{fig:grain8}. For this plot we divide the distance from the star from $2$ to $10\AU$ into 800 equal sized bins and count the number of particles per bin. We can clearly see that 1500 years after we inserted the particles, the distribution clearly deviates from the initial Poisson distribution and instead they clump together. This happens only for particles with a dimensionless  stopping time of the order of unity. 

\begin{figure*}[!t]
\includegraphics[width=\textwidth]{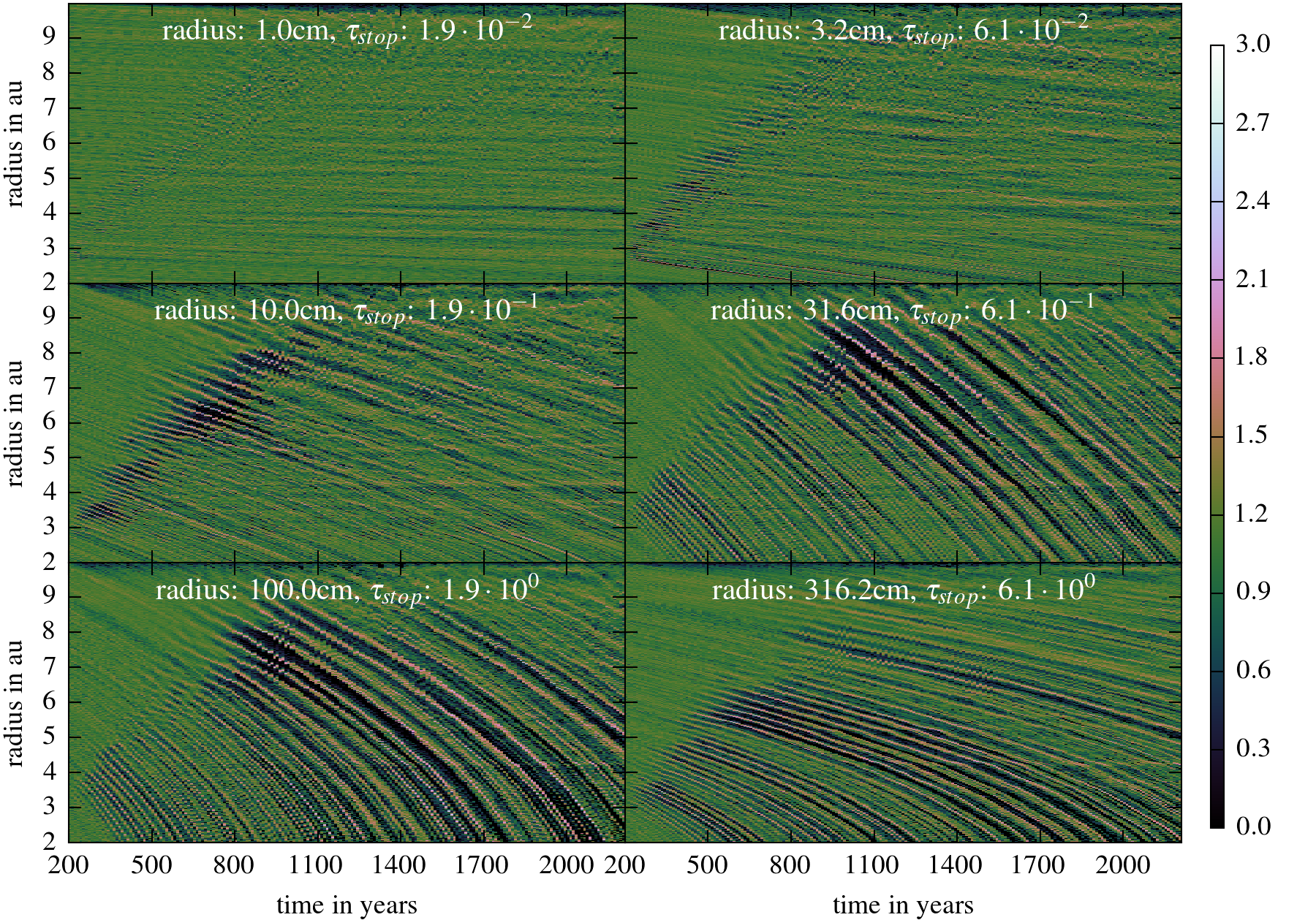}
\caption{Histogram. Visualisation of the radial drift of the particles. We show the logarithm with base 10 of the number of particles per bin. One can see the clumping behaviour as the VSI growths, but also the velocity of the particles. }
\label{fig:mig_pic6}
\end{figure*}

To further illustrate this feature, we show the number of particles per radial bin over time for different particles sizes in Fig. \ref{fig:mig_pic6}. Note that the color scale is logarithmic. For this image and the following analysis we correct the particle density per $1\AU$ to remain constant, as it was at the beginning of the simulations, by weighting the number of particles per bin by the number of particles per $1\AU$. Since the particles move faster in the inner region and the particles that leave the inner region are added in the outer region, we would produce overdensities otherwise. This does not change the bunching statistics, however.

After we insert the particles, the VSI is only active in the inner region, but is quickly spreading out to the whole disc, until after 1000 years the whole domain is active. One can see that the onset of the VSI leads to bunching of the particles.
Due to the bunching we can clearly see the different drift velocities of the particles. But we can also see that at certain radii the particles are sometimes caught in the pressure fluctuations for a short time before moving on, which leads to the visible lines in the image.

\begin{figure}[tb]
\begin{center}
\includegraphics{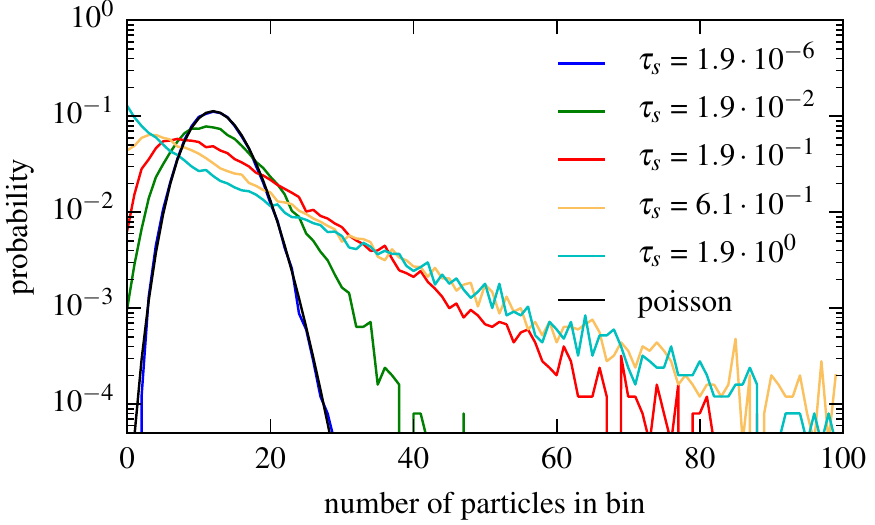}
\end{center}
\caption{Probability to find a certain number of particles in a radial bin with $\Delta r = 0.01\AU$ in the region from 3 to $9\AU$, for different
stopping times (refer to Fig. \ref{fig:grain8}). The average number of particles per bin is 12.5. We averaged over 50 snapshots, each 10 years apart, beginning with year 1500.}
\label{fig:poisson}
\end{figure}

To make the dynamics involved clearer, we calculate another statistic property. This time we count how often a certain number of particles in a radial bin occurs. We average over 50 snapshots, each 10 years apart, beginning 1300 years after we inserted the particles. We then normalise by the total number of particles to find the probability for certain number of particles in a radial bin with width $\Delta r = 0.01$. This is shown in Fig. \ref{fig:poisson}. 

We can see that for the limit of small stopping time we still follow the initial Poisson distribution, which has its peak at 12.5 particles per bin 
(the average number) and decays with $\propto \exp(-n^2)$. These particles are tightly coupled to the gas and can not be caught in pressure fluctuations. In contrast, particles with (dimensionless) stopping time near unity decay with $\propto \exp(-n)$. This increases the likelihood to find bins with a large number of particles which can easily lead to overdensities in dust by a factor of around 10. This is caused by short lived pressure fluctuations originating from the VSI, which briefly slow down the crossing particles. The largest particles again revert to the Poisson statistics, since they are not coupled to the gas.

\subsection{Radial diffusion of particles}

\begin{figure}[tb]
\begin{center}
\includegraphics{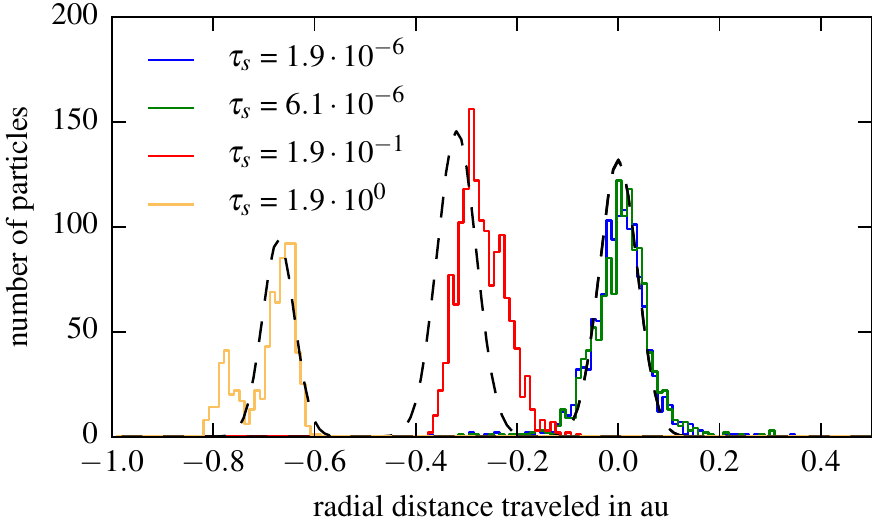}
\end{center}
\caption{Radial particle diffusion over 100 years, after 1000 years. The black dashed lines are calculated from theory for the different stopping times and with $\tau_{eddy} = 0.1$ and $<u_{g,r}^2> = 2 \cdot 10^{-6} \cdot u_{K,1au}^2$, see Eq.~(\ref{eq:radDiff}). }
\label{fig:difRad}
\end{figure}

Next we compare the radial diffusion of the embedded particles with theoretical predictions. Since the power spectra of $u_r,\delta u_\phi,u_\theta$ are similar, the radial diffusion coefficient for particles \citep{Youdin2007Icar..192..588Y} is given by
\begin{equation}
    D_{d,r} = t_{eddy} \frac{\left< u^2_{r}\right> + 4\tau_s^2 \left< \delta u^2_{\phi}\right> + 4\tau_s \left< u_{r} \delta u_{\phi}\right>}{\left(1+\tau_s^2 \right)^2} \,.
    \label{eq:radDiff}
\end{equation}
For the simulation used for Fig. \ref{fig:difRad} we measure $\left<u_{r}^2 \right>_{r,\phi,t} = \left<\delta u_{\phi}^2 \right>_{r,\phi,t} = 2 \cdot 10^{-6} \cdot u_{K,1au}^2$ and $\left< u_{r} \delta u_{\phi}\right>_{r,\phi,t} \leqslant 10^{-8}\cdot u_{K,1au}^2$ near the midplane, indicating isotropic turbulence. With these values and Eq.~(\ref{eq:radialdrift}) for the radial drift we can use the radial diffusion to measure $t_{eddy}$. Surprisingly, from this we calculate a small dimensionless, $\tau_{eddy} =  t_{eddy} \Omega_K$ of $0.1$, compared with the large scale oscillations of the VSI on a timescale of 5 orbits per oscillation. 

In Fig. \ref{fig:difRad} we show the radial distribution of particles with different sizes from which we extracted the radial drift and diffusion during a timespan of 100 years for different particle sizes, again for particles starting at $5\pm 0.5 \AU$. While the smallest particles (blue and green curves) follow directly the prediction, the particles with $\tau_s = 0.19$
(red curve) seem to lag behind the theoretical curve. As described below this difference is caused by the fact that the particles are
more spread out vertically. Even larger particles (yellow) again show bunching behavior and are collected in two distinct peaks, due to being caught in different VSI waves. 

From this we can also calculate the radial Schmidt number, which \citet{Youdin2007Icar..192..588Y} determine for 
homogeneous isotropic turbulence in the $xy$-plane with $\left<u_{r}^2 \right> = \left<\delta u_{\phi}^2 \right>$, to be
\begin{equation}
    \text{Sc}_r = \frac{D_{g,r}}{D_{p,r}} = \frac{\left( 1 + \tau_s^2\right)^2}{1 + 4 \tau_s^2} \,,
    \label{eq:schmidt}
\end{equation}
where for the gas diffusion coefficient, $D_{g,r}$, we use in our case the dust diffusion of the smallest particles.

For smaller particles we find good agreement with Eq.~(\ref{eq:schmidt}) but for $\tau_s$ of unity and larger,
we measure Schmidt numbers smaller as predicted by Eq.~(\ref{eq:schmidt}) by a factor up to three at a stopping time of $\tau_s = 200$.
This is just noticable in Fig. \ref{fig:difRad} for the stopping time of $\tau_s = 1.9$. There we can see that the predicted diffusion fits a single peak well, but not the whole curve, which is 50\% wider. This can be explained by particles crossing large scale VSI modes which are not expected by the theoretical model of homogeneous isotropic turbulence used to calculate the predictions for the Schmidt number.

\subsection{Vertical diffusion of particles}

\begin{figure}[tb]
    \centering
    \includegraphics{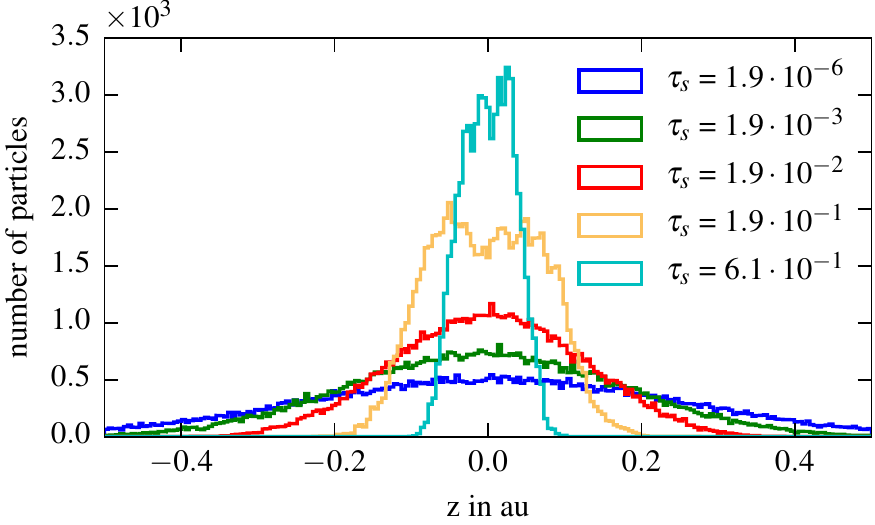}
    \caption{Histogram: Vertical distribution of the particles depending on size, summed over 50 snapshots in the interval from 1200 to 1700 years. Bin size is $\Delta z = 0.02\AU$.}
    \label{fig:diffvert}
\end{figure}

After the radial diffusion shows a small eddy life time, we are now interested in the vertical diffusion.
In Fig. \ref{fig:diffvert} we selected particles between 4.5 and $5.5\AU$ and calculated a histogram of the vertical position to show the vertical distribution of different sized dust particles. For better statistics we added up the 50 last snapshots, spanning from 1200 years to 1700 years. While the small particles with stopping time smaller than unity show the expected Gaussian distribution with a scale height equal to the gas scale height, the particles with stopping time around unity clearly deviate from this. This can be explained by the large scale velocity pattern of the gas with active vertical shear as displayed in Fig. \ref{fig:vx2_rtheta}. The corrugation mode of the gas will move the particles upward away from the disc's midplane, up to the point where the density of the gas is so small that the drag force can no longer overpower gravity. The particle then can swiftly ``surf'' on the updraft of the large scale VSI-mode. Since these corrugation modes oscillate, the particles will also oscillate around the midplane.

For isotropic turbulence \citet{Dubrulle1995Icar..114..237D} and \citet{Zhu2015ApJ...801...81Z} calculate a dust scale height of
\begin{equation}
    h_p = \frac{h}{\sqrt{H^2 \Omega^2 t_s/ \left(  \left< u^2_{z} \right> t_{eddy} \right)+1}} \,,
    \label{eq:hdust}
\end{equation}
where $h=H/R$ is the relative gas scale height. 
Together with the measured velocity dispersion $<u_{z}^2> = 5 \cdot 10^{-6} \cdot u_{K,1au}^2$ we can use this equation to calculate again the eddy timescale by fitting a Gaussian to the data for different stopping times, thus extracting the dust scale height. Note that for a very small stopping time this does not work, since the dust scale height is equal to the gas scale height, independent of the eddy time scale. Also we superimpose two Gaussian with the same scale height for the distributions with two peaks, which then fits well. These peaks are then usually two scale heights apart for particles with dimensionless stopping time around one. Without this scheme, we would get a larger scale height and thus a larger eddy lifetime, which makes sense, since they are caused by the large scale structures that have a long life time.  We present the results in table \ref{tab:difvert}. Again we find $\tau_{eddy} \approx 0.2$, which is similar for other types of turbulence, for example MRI has $\tau_{eddy} \approx 1$  \citep{Youdin2007Icar..192..588Y,Carballido2011MNRAS.415...93C}. Note that the larger eddylife times for the larger particles indicate that the eddylife time is large in the midplane, since the smaller particles also see the eddylife time of the corona. 

\begin{table}[tb]
\begin{tabular}{cccc}
\hline \hline
radius & stopping time $\tau_s$ & $h_{{p,z}}$ & $\tau_{{eddy,z}}$ \\
cm & at 5\AU & $h_{gas}$ & at 5\AU \\
\hline
$  1.0 \cdot 10^{ -4 }$ & $  1.9 \cdot 10^{ -6 }$ & $  1.0 \cdot 10^{ 0 }$ & - \\
$  3.2 \cdot 10^{ -4 }$ & $  6.1 \cdot 10^{ -6 }$ & $  9.9 \cdot 10^{ -1 }$ & - \\
$  1.0 \cdot 10^{ -3 }$ & $  1.9 \cdot 10^{ -5 }$ & $  9.3 \cdot 10^{ -1 }$ & - \\
$  3.2 \cdot 10^{ -3 }$ & $  6.1 \cdot 10^{ -5 }$ & $  8.5 \cdot 10^{ -1 }$ & - \\
$  1.0 \cdot 10^{ -2 }$ & $  1.9 \cdot 10^{ -4 }$ & $  8.3 \cdot 10^{ -1 }$ & - \\
$  3.2 \cdot 10^{ -2 }$ & $  6.1 \cdot 10^{ -4 }$ & $  7.6 \cdot 10^{ -1 }$ & $  8.1 \cdot 10^{ -2 }$ \\
$  1.0 \cdot 10^{ -1 }$ & $  1.9 \cdot 10^{ -3 }$ & $  6.6 \cdot 10^{ -1 }$ & $  1.4 \cdot 10^{ -1 }$ \\
$  3.2 \cdot 10^{ -1 }$ & $  6.1 \cdot 10^{ -3 }$ & $  5.1 \cdot 10^{ -1 }$ & $  2.1 \cdot 10^{ -1 }$ \\
$  1.0 \cdot 10^{ 0 }$ & $  1.9 \cdot 10^{ -2 }$ & $  4.3 \cdot 10^{ -1 }$ & $  4.3 \cdot 10^{ -1 }$ \\
$  3.2 \cdot 10^{ 0 }$ & $  6.1 \cdot 10^{ -2 }$ & $  3.0 \cdot 10^{ -1 }$ & $  6.1 \cdot 10^{ -1 }$ \\
$  1.0 \cdot 10^{ 1 }$ & $  1.9 \cdot 10^{ -1 }$ & $  1.9 \cdot 10^{ -1 }$ & $  7.0 \cdot 10^{ -1 }$ \\
$  3.2 \cdot 10^{ 1 }$ & $  6.1 \cdot 10^{ -1 }$ & $  9.4 \cdot 10^{ -2 }$ & $  5.4 \cdot 10^{ -1 }$ \\
$  1.0 \cdot 10^{ 2 }$ & $  1.9 \cdot 10^{ 0 }$ & $  3.5 \cdot 10^{ -2 }$ & $  2.3 \cdot 10^{ -1 }$ \\
$  3.2 \cdot 10^{ 2 }$ & $  6.1 \cdot 10^{ 0 }$ & $  1.8 \cdot 10^{ -2 }$ & $  1.9 \cdot 10^{ -1 }$ \\
$  1.0 \cdot 10^{ 3 }$ & $  1.9 \cdot 10^{ 1 }$ & $  6.7 \cdot 10^{ -3 }$ & $  8.6 \cdot 10^{ -2 }$ \\
\hline
\end{tabular}
\caption{Measured dust scale heights and inferred eddy lifetimes, from Eq.~(\ref{eq:hdust}), for the inviscid isothermal model. A Gaussian $f(z) = N_0 \exp{(-(z\pm\mu)^2/(2 r^2 h_{p}^2 })$ was fitted to the data of the vertical distribution to find $h_{p}$ and calculated $\tau_{eddy}$. The first five values for $\tau_{eddy}$ were not calculated, since the gas scale height is nearly equal to the dust scale height.} \label{tab:difvert}
\end{table}

\subsection{Collision statistics}

In this section we evaluate the relative velocity distribution for colliding particles. Since we do not have enough particles to directly measure this distribution, we take each particle between 4 and $7\AU$ and check for other particles in a sphere with radius smaller than $0.05\AU$. To make them independent of translation we remove the Kepler velocity (for independence of radius) and rotate all particles into the same $r$-$\theta$ plane, after we calculate the distance (for independence of the azimuthal angle, since they are all in circular orbits).  We then calculate histograms of the relative velocity between two particles depending on the size of the particles. We normalise by the number of particles to get a probability.  

\begin{figure}[tb]
    \centering
    \includegraphics{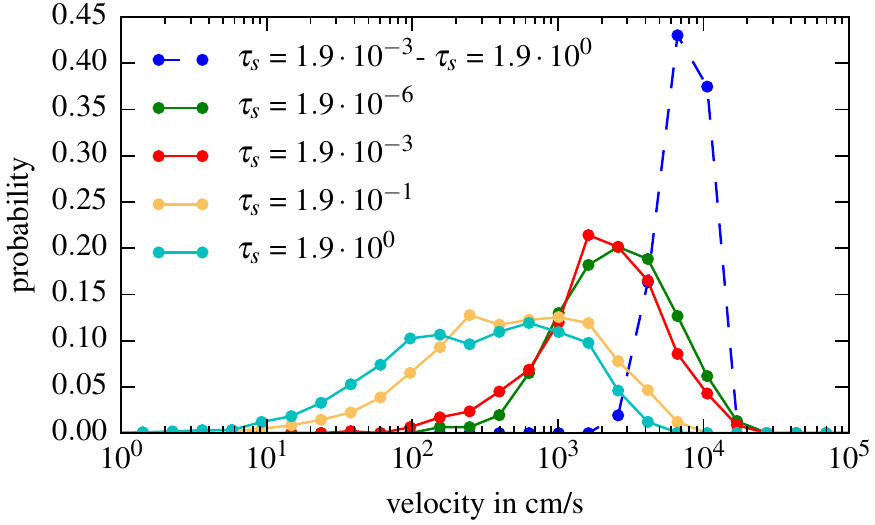}
    \caption{Histogram: Relative velocity between particles for different stopping times after 2200 years. The dashed line represents particles of different sizes and the solid lines denote collisions between same sized particles. Only particles with separation smaller than $0.05\AU$ are considered for potential collisions. Lines to guide the eye.}
    \label{fig:collvel}
\end{figure}

This can be seen in Fig. \ref{fig:collvel}. Lost by the normalization procedure is the fact that there are around 300  same sized particles in the sphere for small particles, while there are around 11000 same sized particles with stopping time around unity in the sphere. Since the median relative velocity for larger particles is reduced by an order of magnitude relative to the smaller particles, the number of collisions will also be reduced. The blue dashed line represents the relative velocity between small and large particles, since those are dominated by the velocity difference that is created by the strongly coupled small particles, moving with the gas, and weakly coupled particles moving with Keplerian velocity.

\begin{figure}[tb]
    \centering
    \includegraphics{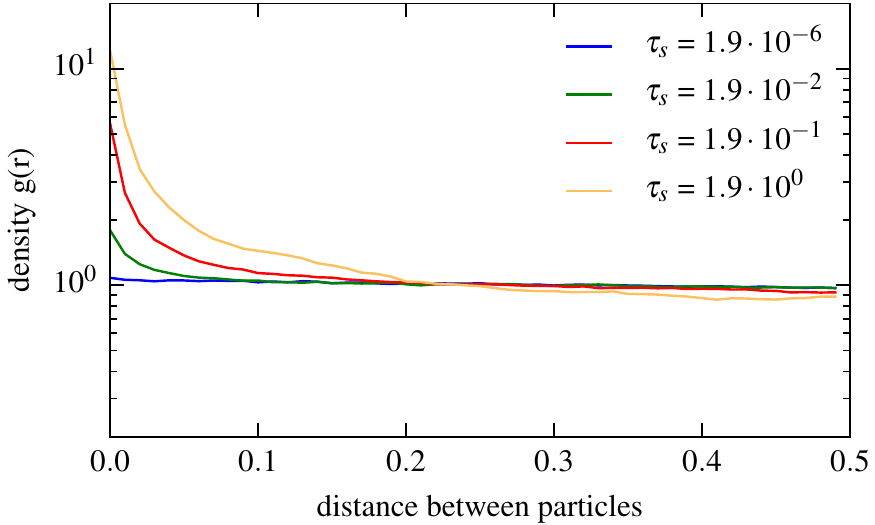}
    \caption{Pair correlation function: enhancement of surface density in a shell with radius $r$ for different stopping times. Averaged over 5 snapshots  in the interval from 1800 to 2200 years.}
    \label{fig:corr}
\end{figure}

Using an alternative way of estimating the bunching behaviour of particles, we
also calculate the pair correlation function $g(r)$. 
\begin{equation}
    g(r) = \frac{V}{\pi r \Delta r N^2} \sum_i^N \sum_{j \neq i}^N \delta(r - d_{ij})
    \label{pariCor}
\end{equation}
where $V$ is the area of integration, $d_{ij}$ is the distance between particle i and j and $\delta(r)$ is one if $|r|<\Delta r/2$ and we use $\Delta r = 0.01\AU$.
This function returns one for Poisson distributed particles and larger than one if there is an increased surface density in the ring around the particles at this radius and thus picks up 2D clustering instead of the 1D clustering in the earlier analysis. We calculate this property for particles between $4$ and $7\AU$ projected in the $r-\phi$ plane, and
show it for different particle sizes in Fig. \ref{fig:corr}. We can see that small particle positions are uncorrelated, but the particle positions with stopping time near unity display a clear correlation, as we could already infer from Fig. \ref{fig:poisson}. Particles with a stopping time closer to unity have a larger correlation length. This makes the VSI a possible candidate to trigger the streaming instability.

\section{3D-simulations: viscosity}\label{sec:viscosity}
In this section we present the same simulation but using now a dimensionless kinematic viscosity coefficient of $\nu=5 \cdot 10^{-7}$, in order to check the influence of viscosity, which in turn influences the $\alpha$-parameter and the velocity dispersion.  We will also add the same amount of viscosity to the model with radiation transport, as shown below. This viscosity corresponds to an $\alpha$-value of $9\cdot 10^{-5}$ at 5 au or $4\cdot 10^{-5}$ at 25 au. There it will limit the smallest length scale of the VSI, which can not be resolved otherwise. We only show the results if there is a clear difference to the previous simulation. From the hydro-dynamic perspective they are very similar, but the $\alpha$-parameter is smaller by a factor of 2 and the wavelength of the instability slightly smaller.

\subsection{Radial drift}  \label{sec:raddif}
\begin{figure}[bt]
\begin{center}
\includegraphics{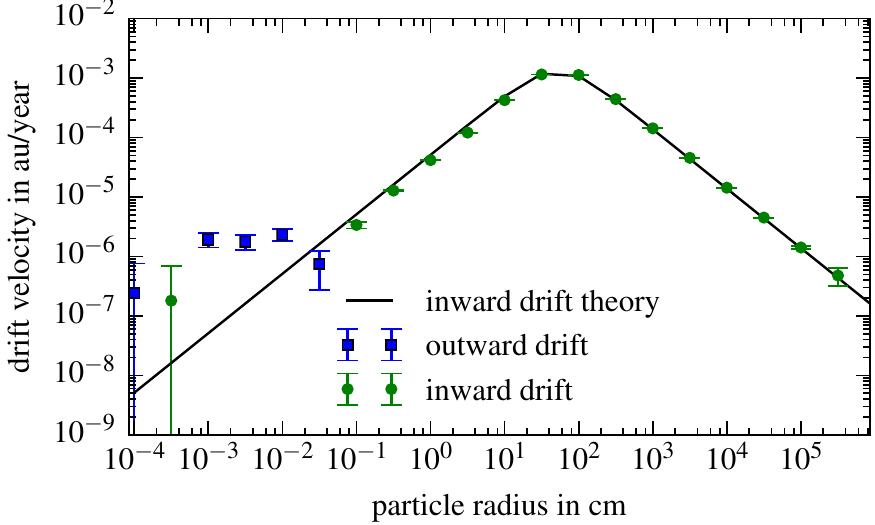}
\caption{The drift velocity of the dust particles depending on the radius of the particles at $r=5\AU$ for a viscous disc. This is compared to expected transport for pressure supported discs. Simulation with resolution $1024\times256\times64$ and viscosity $\nu=5 \cdot 10^{-7}$. Different colors are used for inward and outward drift. We estimate the error from the radial diffusion coefficent.}
\label{fig:migration5}
\end{center}
\end{figure}

If we include viscosity and repeat the analysis for radial drift, here from 2700 years to 3700 years, due to the slower diffusion, we can see in Fig. \ref{fig:migration5} 
that particles larger than about 0.1 cm drift inward with approximately the theoretical speed (see Eq.~\ref{eq:radialdrift}) with slightly larger deviations than in the inviscid case.
The smallest particles now clearly drift away from the star at $r=5\AU$. This is true for different radii of the disc. 

\begin{figure}[bt]
\begin{center}
\includegraphics{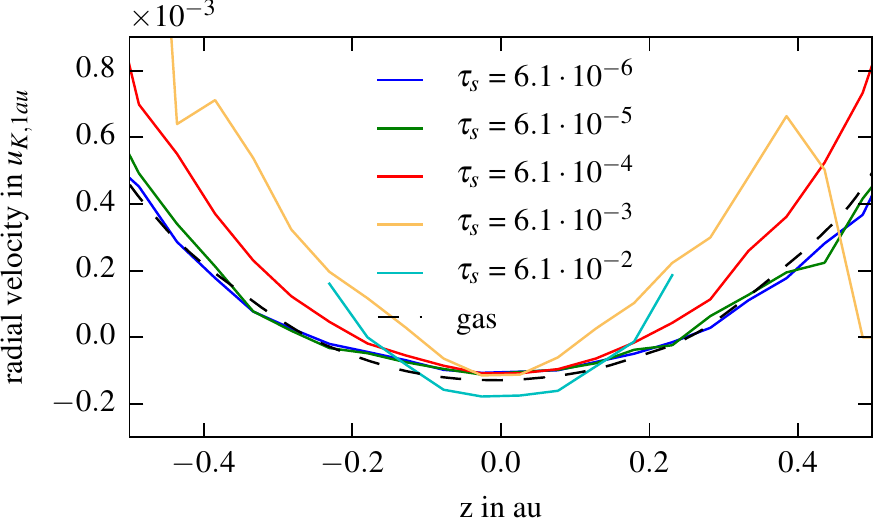}
\end{center}
\caption{The drift velocity of the dust particles at $r=5\AU$ depending on the vertical direction compared to the gas velocity for the simulation with viscosity $\nu=5 \cdot 10^{-7}$ averaged from 2700 to 3700 years.}
\label{fig:drift5}
\end{figure}

\begin{figure}[bt]
\begin{center}
\includegraphics{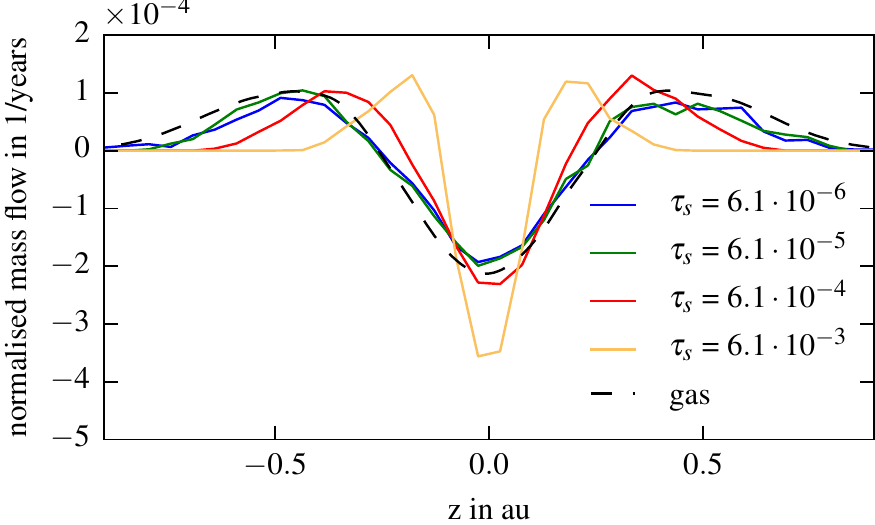}
\end{center}
\caption{The normalised mass flux $u_{\text{drift}} \rho /\Sigma$ at $r = 5 \AU$ for the simulation with viscosity $\nu=5 \cdot 10^{-7}$ averaged from 2700 to 3700 years.}
\label{fig:massflow}
\end{figure}

In Fig. \ref{fig:drift5} we see that the particles are moving inwards at the midplane and outwards otherwise. 
The smallest particles follow the gas velocity and  larger particles are moving outwards away from the midplane faster than the gas, similar to \citet{Takeuchi2002ApJ...581.1344T} where they move outwards even though the gas is moving inwards, due to the gas being super-keplerian in the disc's corona. This leads to a massflow (shown in Fig. \ref{fig:massflow}) that is inward in the midplane and outward farther away from the midplane. 
As seen in Figs.~\ref{fig:drift5} and \ref{fig:massflow} small particles (blue and green line) show the same radial velocity profile and mass flux behaviour as the disk's gas flow (black dashed lines). To estimate the overall mean flow of the particles it is necessary to consider the 
vertical dependence of the mass flow (Fig.~\ref{fig:massflow}) rather than the radial velocity $z$-profile (Fig.~\ref{fig:drift5}).
For example, from Fig.~\ref{fig:drift5} we notice that the particles with stopping time $\tau = 6.1 \cdot 10^{-5}$ (green line, $3.2 \cdot 10^{-3}$ cm radius)
move inward slower than the particles with stopping time $\tau = 6.1 \cdot 10^{-4}$ (red line,  $3.2 \cdot 10^{-2}$ cm radius)
even though the mean radial dust velocities in Fig. \ref{fig:migration5} are {\itshape smaller} for the larger particles. This is due to the different vertical distribution of the particles (see Fig. \ref{fig:massflow} and table \ref{tab:verdif5}). In addition, the velocity and mass flow profiles of 
the smallest displayed particles (blue lines in Figs.~\ref{fig:drift5} and \ref{fig:massflow}, $3.2 \cdot 10^{-4}$ cm radius) 
look very similar to the particles with stopping time $\tau = 6.1 \cdot 10^{-5}$ (green line), while having drift rates in the opposite direction
in Fig. \ref{fig:migration5}.

Thus the net particle flow is very sensitive to the gas flow profile and the ratio of particles near the midplane, which is decided by the stopping time. This also means that, independent of the mean flow, there will always be a small fraction of particles drifting away from the star, faster than one would expect from diffusion alone.

\subsection{Radial diffusion} 
\begin{figure}[t]
    \centering
    \includegraphics{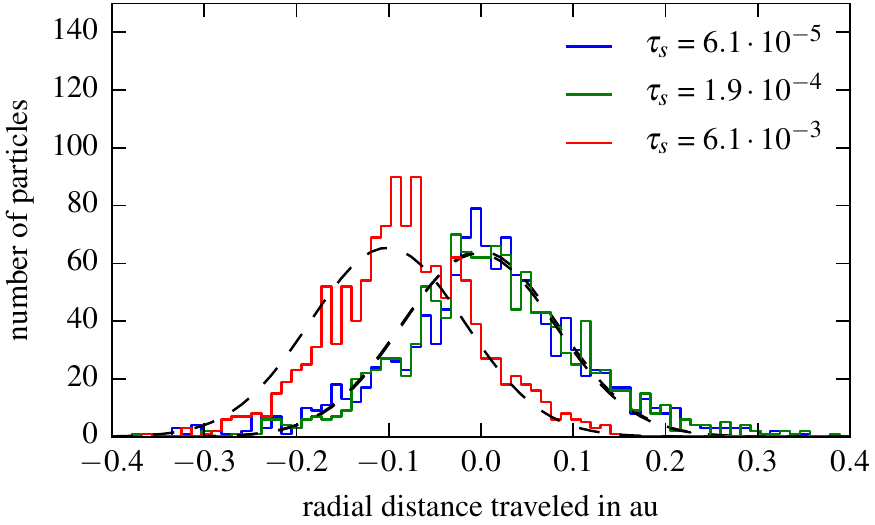}
    \caption{Radial diffusion over 1000 years after 2700 years for the simulation with viscosity $\nu = 5 \cdot 10^{-7}$. The dashed lines are calculated from theory for the different stopping times and with the same $\tau_{eddy} = 0.1$ and $<u_{gas,radial}^2> = 1 \cdot 10^{-6} \cdot u_{Kepler,1au}^2$. Compare with figure \ref{fig:difRad}, but note that here the particles had 10 times as much time to diffuse.}
    \label{fig:difRadeta5alt}
\end{figure}

The increase in viscosity leads to an decrease of the velocity dispersion to $<u_{r}^2> = 1 \cdot 10^{-6} \cdot u_{Kepler,1au}^2$ which is smaller by a factor of two. In Fig. \ref{fig:difRadeta5alt} we compare our results to theoretical predictions with $\tau_{eddy}=0.1$. The diffusion is clearly smaller than in the inviscid case, as predicted by the decreased velocity dispersion. This is in contrast to the vertical eddy lifetime were we measure an increase in eddy lifetime due to vertical diffusion. 

We can also see (right wing of green and blue curves) that a small fraction of the small particles diffuse faster outwards than the rest of the particles. These are particles far away from the midplane, where the stopping time is magnitudes larger, typically larger than $10^{-2}$, and the gas flow is in average outwards. These weaker coupled particles can quickly travel a short distance away from the star, before drifting back inwards nearer to the midplane. The difference in radial drift between theory and simulation noticed in Fig.~\ref{fig:migration5} causes the offset for the results of the longer stopping time (red curve).

\subsection{Vertical Diffusion}
For vertical diffusion we obtain slightly larger eddy lifetimes as can be seen in table \ref{tab:verdif5} where we measured from 2700 years to 3700 years.
In these simulations it took four times as long for the dust scale height to converge to the gas scale height for the smallest
particles, even though the vertical velocity dispersion is identical to the inviscid case.

\begin{table}
\begin{tabular}{cccc}
\hline \hline
radius & stopping time $\tau_s$& $h_{{p,z}}$ & $\tau_{{eddy,z}}$ \\
cm & at 5\AU & $h_{{gas}}$ & at 5\AU \\
\hline
$  1.0 \cdot 10^{ -4 }$ & $  1.9 \cdot 10^{ -6 }$ & $  9.5 \cdot 10^{ -1 }$ & - \\
$  3.2 \cdot 10^{ -4 }$ & $  6.1 \cdot 10^{ -6 }$ & $  9.4 \cdot 10^{ -1 }$ & - \\
$  1.0 \cdot 10^{ -3 }$ & $  1.9 \cdot 10^{ -5 }$ & $  1.0 \cdot 10^{ 0 }$ & - \\
$  3.2 \cdot 10^{ -3 }$ & $  6.1 \cdot 10^{ -5 }$ & $  9.5 \cdot 10^{ -1 }$ & - \\
$  1.0 \cdot 10^{ -2 }$ & $  1.9 \cdot 10^{ -4 }$ & $  8.9 \cdot 10^{ -1 }$ & - \\
$  3.2 \cdot 10^{ -2 }$ & $  6.1 \cdot 10^{ -4 }$ & $  8.0 \cdot 10^{ -1 }$ & $  1.0 \cdot 10^{ -1 }$ \\
$  1.0 \cdot 10^{ -1 }$ & $  1.9 \cdot 10^{ -3 }$ & $  5.5 \cdot 10^{ -1 }$ & $  8.1 \cdot 10^{ -2 }$ \\
$  3.2 \cdot 10^{ -1 }$ & $  6.1 \cdot 10^{ -3 }$ & $  3.9 \cdot 10^{ -1 }$ & $  1.1 \cdot 10^{ -1 }$ \\
$  1.0 \cdot 10^{ 0 }$ & $  1.9 \cdot 10^{ -2 }$ & $  3.3 \cdot 10^{ -1 }$ & $  2.2 \cdot 10^{ -1 }$ \\
$  3.2 \cdot 10^{ 0 }$ & $  6.1 \cdot 10^{ -2 }$ & $  2.8 \cdot 10^{ -1 }$ & $  4.8 \cdot 10^{ -1 }$ \\
$  1.0 \cdot 10^{ 1 }$ & $  1.9 \cdot 10^{ -1 }$ & $  2.1 \cdot 10^{ -1 }$ & $  8.3 \cdot 10^{ -1 }$ \\
$  3.2 \cdot 10^{ 1 }$ & $  6.1 \cdot 10^{ -1 }$ & $  9.4 \cdot 10^{ -2 }$ & $  5.2 \cdot 10^{ -1 }$ \\

\hline
\end{tabular}
\caption{Measured dust scale heights and inferred eddy lifetimes, from Eq.~(\ref{eq:hdust}),  for the viscous isothermal model. 
 We fitted a Gaussian $f(z) = N_0 \exp{(-(z\pm\mu)^2/(2 r^2 h_{p}^2 }))$ to the data of the vertical distribution to find $h_{p}$ and calculated $\tau_{eddy}$ for the simulation with viscosity of $5\cdot 10^{-7}$. }
\label{tab:verdif5}
\end{table}

\subsection{Clustering}
\begin{figure}[tb]
\includegraphics{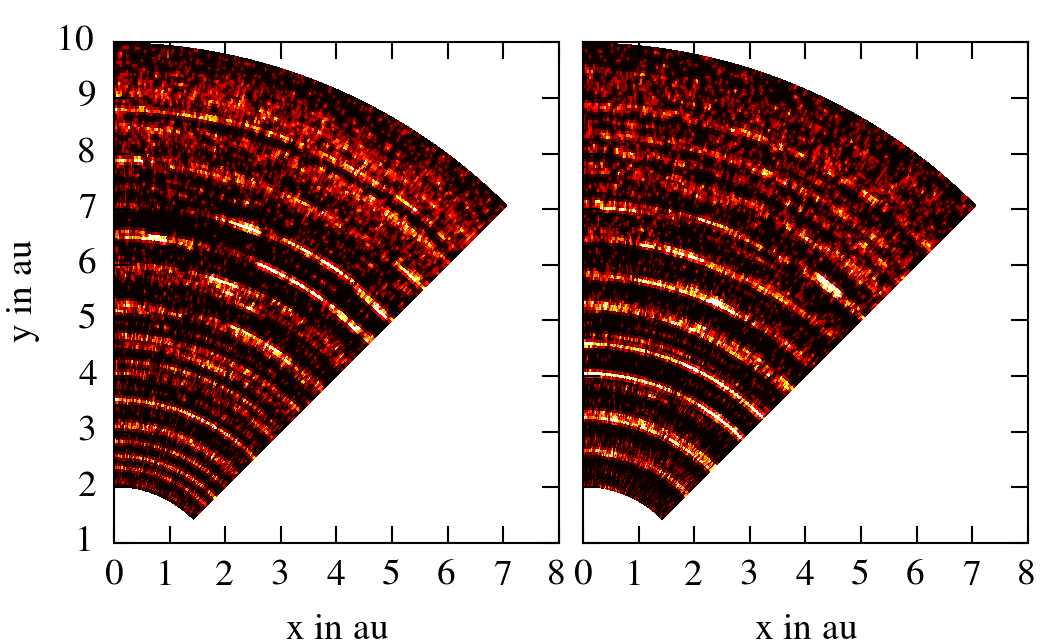}
\caption{Dust distribution for the disc with viscosity of $5\cdot 10^{-7}$. Shown are the particles with dimensionless stopping time $\tau_s = 0.6$
 (left panel) and $\tau_s = 1.9$ (right panel) at $5\AU$.}
\label{fig:eta5_rphi}
\end{figure}

\begin{figure}[t]
    \centering
    \includegraphics{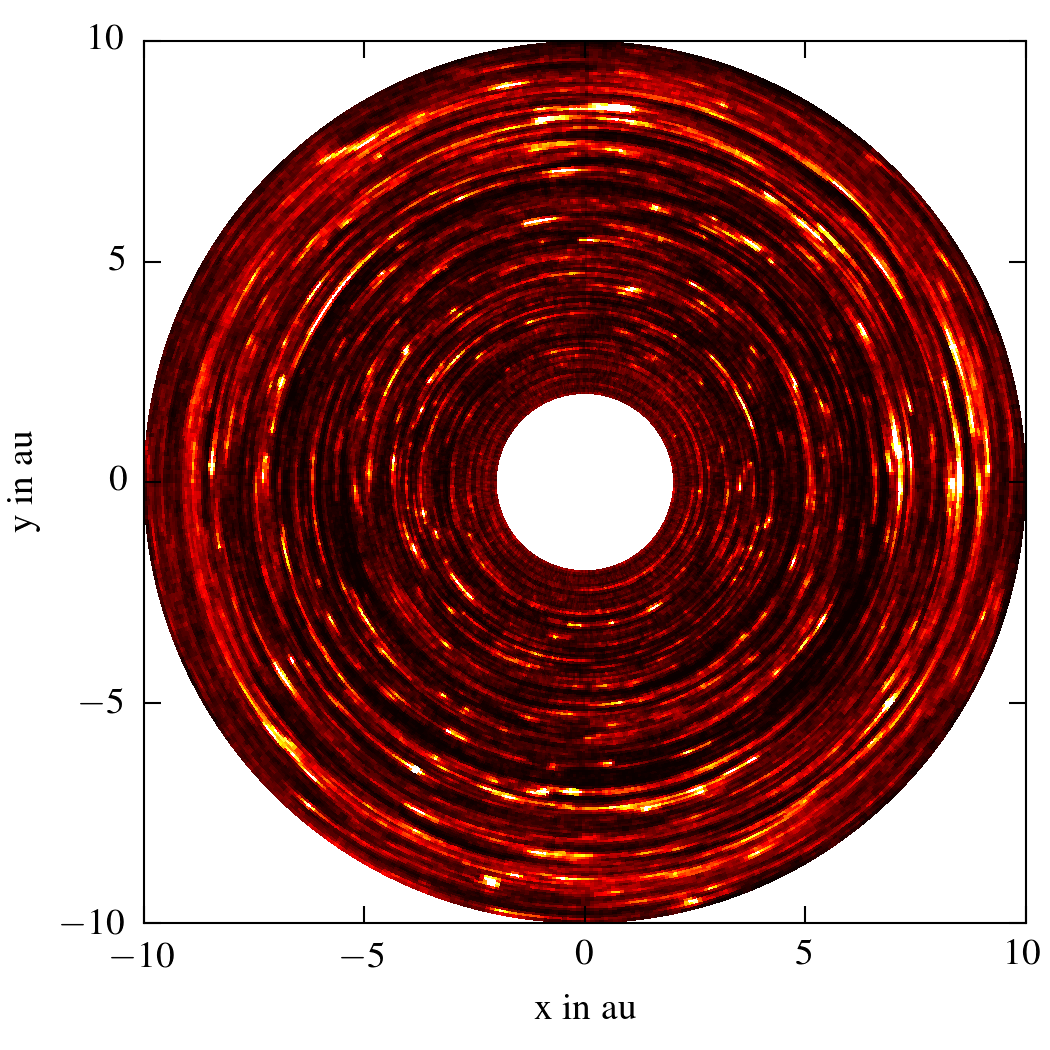}
    \caption{Dust distribution for the full disc with viscosity of $10^{-7}$. Shown are the particles with dimensionless stopping time  $\tau_s = 0.6$. For this simulation we added 500,000 particles of the same size after 3400 years and let them evolve for 400 years. }
    \label{fig:eta1_rphi}
\end{figure}

Finally we present the particle distribution in the $r-\phi$ plane in Fig. \ref{fig:eta5_rphi}. We show the particles with dimensionless stopping time of $\tau_s = 0.6$ and $\tau_s = 1.9$ at $5\AU$. One can see for both displayed particles sizes that the VSI modes have produced nearly axisymmetric clusters. The bunching leading to the ring structure is strongest for the simulation with high viscosity. We also show in Fig. \ref{fig:eta1_rphi} the same effect for the full disc with small viscosity of $10^{-7}$ and resolution of $512\times128\times512$. For this simulation we also increased the number of particles to 500,000. We can see that the ring structure already seen in the histogram of Fig. \ref{fig:grain8} does indeed persist even in a full disc. Particles with more than a magnitude larger or smaller stopping time do not show this features. 

\section{3D-simulations: radiative model}\label{radiative}
In this section we present the results for a radiative disc with radiation transport and irradiation from the central star.
In contrast to our first paper \citep{Stoll2014A&A...572A..77S}, we model the stellar irradiation in a more realistic fashion as coming from the central star, similar to the treatment in \citet{2013A&A...549A.124B}.
This star has a temperature of \unit[4000]{K} and a radius of \unit[4]{$R_{\odot}$}.
\subsection{Setup}
For this simulation we have to take into account that the cooling time has to be sufficiently small for the VSI to be active \citep{Nelson2013MNRAS.435.2610N,Lin2015ApJ...811...17L},
which made changes in the domain necessary. We moved the radial extent of the disc for the inner boundary from 2 to $8\AU$ and for the outer boundary from 10 to $80\AU$. This radial range is expected to be the  active region of the VSI \citep{Lin2015ApJ...811...17L}. 
We simulate again one eighth of the disc in the azimuthal direction and this domain is resolved by $1200\times 260 \times 60$ grid cells.
We also changed the density profile exponent from $p = -1.5$ to a value that is more in line with the observations $p = -1.8$ \citep{2011ARA&A..49...67W}.
Initially the temperature drops with $T = T_0 \cdot r^{-1}/r_0$, thus we have $\rho = \unitfrac[10^{-9} ]{g}{cm^3} \cdot r^{-1.8}/r_0$.
This translates to $\Sigma = \unitfrac[1700]{g}{cm^2} $ at $1\AU$, which corresponds to is the MMSN-model with a shallower decay of the density.
The radiation transport then quickly leads to a new equilibrium with $T = \unit[900]{K} \cdot r^{-0.6}/r_0$ in the corona and $T = \unit[700]{K} \cdot r^{q}/r_0$ in the midplane, where the temperature gradient exponent $q$ varies slightly around the mean of $q=-0.9$, from  $q = -1.1$ in the inner region to $q=-0.6$ in the outer region. 

During the evolution to the new equilibrium we damp the velocities in the whole disc.

We add a small viscosity of $\nu = 5 \cdot 10^{-7}$. This suppresses the VSI in the inner region, where it would otherwise be weakly active, but due to a wavelength on grid scale clearly not resolved, which in turn would lead to unphysical numerical artifacts. As shown above, we observe only a small change of the VSI activity in the active domain with viscosity enabled compared to the inviscid case. Thus we see no harm in adding it.

In our first paper on the behaviour of the VSI in radiative discs we considered only vertical irradiation onto the disc surfaces \citep{Stoll2014A&A...572A..77S}.
Here, we make the simulations more realistic and irradiate the disc from a central stellar source from the origin along the radial direction, see \citet{2013A&A...549A.124B}.
In this procedure the inner rim of the disc in our simulation is directly exposed to the stellar irradiation. To prevent unphysical heating of the midplane at the inner boundary, we absorb the irradiation flux coming from the star in a fictitious ghost cells with a width 0.25\AU\, using the gas properties of the adjacent innermost active cells
of the domain.

\begin{figure}[t]
    \centering
    \includegraphics{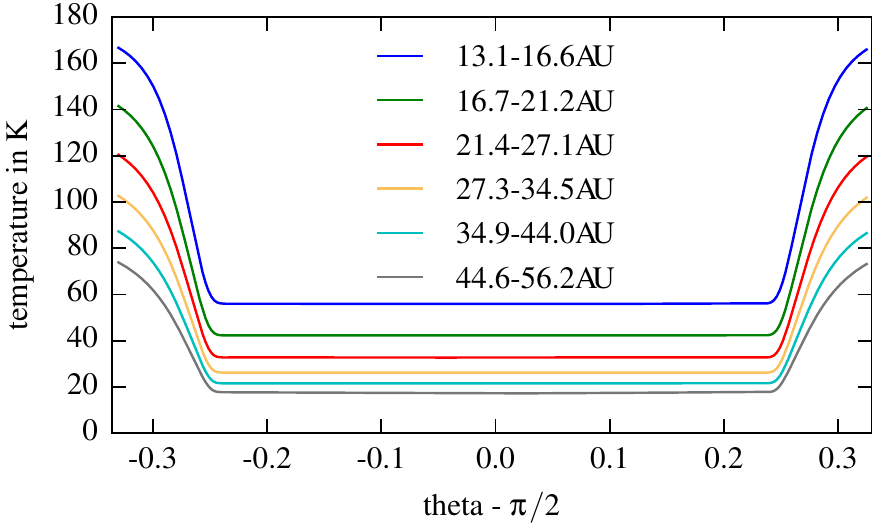}
    \caption{The vertical temperature profile for the irradiated disc in the quasi-equilibrium state at different distances from the central star.}
    \label{fig:irr_tempz}
\end{figure}

For the irradiation opacity we choose a value 10 times higher than the gas opacity, to compensate for the fact, that this radiation is emitted by a hot star and not the surrounding gas. This leads to a heated corona with a cooler midplane as can be seen in Fig. \ref{fig:irr_tempz}, instead of the cooler corona in \citet{Stoll2014A&A...572A..77S}. At the boundary of the corona we can also see a change in the VSI mode. They have a larger wavelength in the hotter corona region and split where the temperature changes to a smaller wavelength in the midplane, see lower panel Fig.~\ref{fig:irr_cooling}.

To make a direct comparision between the isothermal and radiative case, we ran an additional isothermal model with $p=-1.8$ and $q=-0.9$ and $\nu = 5 \cdot 10^{-5}$ and damping in the vertical and radial velocity in the region between 8 and 10\AU\ to avoid boundary effects.  In principle one could also compare the radiative case directly to the isothermal models from section \ref{isothermal}, because in isothermal simulations the unit of length is not fixed and can be scaled to a different regime. However, the gradients in density and temperature are not the same. We thus included a new isothermal model that can be directly compared to.

\subsection{Hydrodynamic properties}

\begin{figure}[t]
    \centering
    \includegraphics{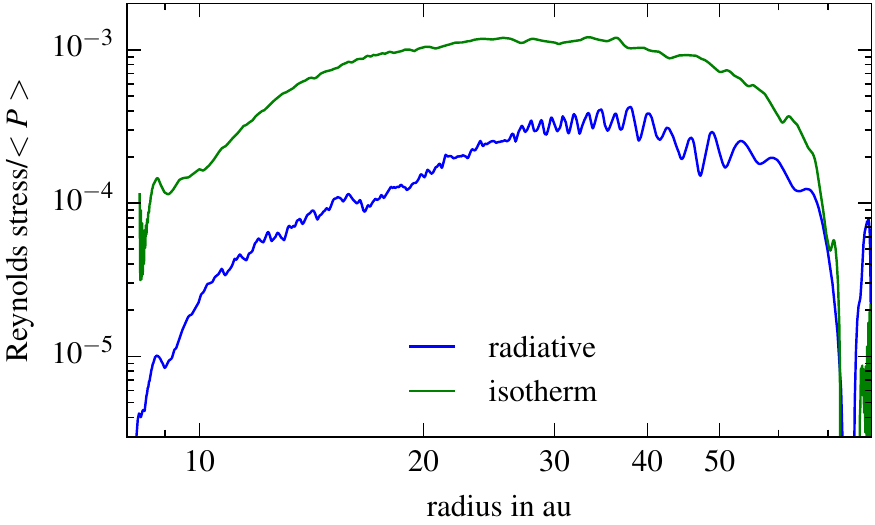}
    \caption{The $\alpha$-parameter for the irradiated disc, calculated with time averaging. We average from 7500 years to 37500 years using 60 snapshots. }
    \label{fig:irr_alpha}
\end{figure}

\begin{figure}[t]
    \centering
    \includegraphics{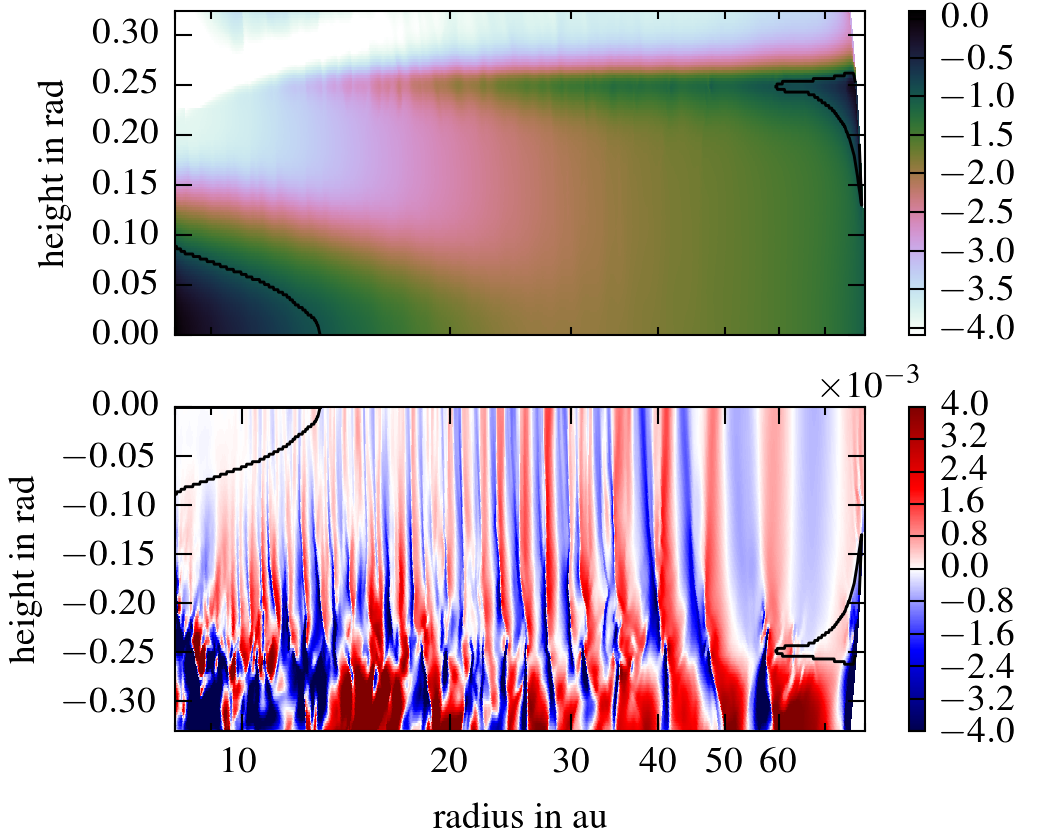}
    \caption{The dimensionless cooling time (upper panel) and vertical velocity (lower panel) for the irradiated disc after 13500 years. The top panel shows the
     upper half of the disc while the lower panel the lower one at the same time slice. 
     The black line indicates the location of the critical cooling time $\tau_{crit}$ (see text), which separates the active from the inactive region. }
    \label{fig:irr_cooling}
\end{figure}

We begin by presenting the $\alpha$-parameter in Fig. \ref{fig:irr_alpha}, here calculated by time averaging the azimuthal velocity, since the equilibrium velocity cannot be computed analytically for radiative discs with a vertically varying temperature. In the inner region the VSI is suppressed by the viscosity of $\nu = 5 \cdot 10^{-7}$ on small wavelengths and by the high cooling time on large wavelengths. Compare this to the isothermal simulation, where the same viscosity is not able to suppress even at $4\AU$ (see Fig. \ref{fig:alpha_iso}). This is followed by an active region beginning at $15\AU$ where we reach $\alpha = (1-4) \cdot 10^{-4}$, which is still smaller than the isothermal simulations. The drop off in the outer region may be linked to the reduced activity in this region, see also Fig. \ref{fig:irr_cooling}, but is also visible in the isothermal model. 

As the VSI is critically dependent on small cooling times, we analyse the cooling times due to radiative diffusion in the irradiated disc models. The radiative diffusion coefficient is given by:
\begin{equation}
    \eta_{rad} = \frac{4 \lambda a c T^3}{\kappa_R \rho^2 c_v}
\end{equation}
where $\lambda$ is the flux limiter, $a$ the radiation constant, $c$ the speed of light and $\kappa_R$ the Rosseland mean opacity. To calculate the cooling time we also need the appropriate length scale. For the optically thick region we simply take the length scale of the perturbation, which we approximate as a fourth of the scale height $l_{thick} = H/4$. In the optically thin region we use the optical mean free path $l_{thin} = 1/\kappa_R \rho$ for the length scale. This leads to a combined dimensionless cooling time of
\begin{equation}
    \tau_{cool} = \frac{l_{thick}^2 + l_{thin}^2}{\eta_{rad}} \Omega_K  \,.
    \label{}
\end{equation}
In Fig. \ref{fig:irr_cooling} we compare the cooling time, $\tau_{cool}$, as calculated from our numerical irradiated disc models with the critical cooling time,
$\tau_{crit}$, as estimated by \citet{Lin2015ApJ...811...17L}, who compared the destabilising vertical shear rates with the stabilising  vertical buoyancy frequency. 
They obtained
\begin{equation}
    \tau_{crit} = \frac{h |q|}{1-\gamma} \,.
    \label{<+label+>}
\end{equation}
We see a good agreement in the inner region between the active regions as predicted by the critical cooling time and the active regions in our simulation. The inner midplane region up to $10\AU$ is completely inactive and the following region which is also predicted to be inactive is only active with a higher order mode. Note that without viscosity one expects modes with higher wavenumber in this region. 

In the outer region beyond $60\AU$ the VSI is inactive despite a small enough cooling time. 
This may be due the dynamics of the VSI that shows larger wavelengths in the outer region, thus requiring a smaller cooling time, or to boundary effects.

One can also see that the jump in temperature and cooling time, that also defines the boundary between disc and corona, creates a boundary for the VSI, where the surface modes can attach to \citep{Barker2015MNRAS.450...21B}.

\subsection{Dust properties}
\begin{figure}[t]
    \centering
    \includegraphics{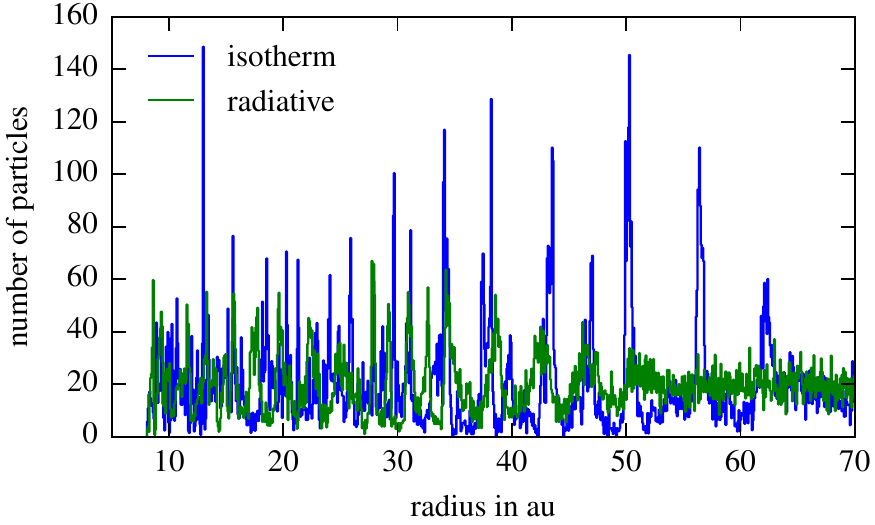}
    \caption{Histogram of particles with size of 31cm and stopping time $\tau_s = 0.2$, after 13500 years. We divide the radial domain from $8$ to $80\AU$ into 1000 bins and count the number of particles in each bin. The average number of particles per bin is 20.0.}
    \label{fig:grain11}
\end{figure}

For this simulation we add 20,000 particles per size after 1000 years. In Fig. \ref{fig:grain11} we present a histogram of the distribution of particles with a size of 31cm, after 13500 years. We see that in the outer region with the inactive VSI the particles are still Poisson distributed, but in the active region they are caught in the eddies. The particles in the outer region are only collected weakly, since the VSI is reduced, due to the large cooling time. The isothermal case we compare to has clustering throughout the whole disc and the overdensities are stronger by a factor of around two, even though the velocity dispersion is higher by a factor of 5 to 10.

These results show that even with realistic cooling times, the VSI can create small axisymmetric regions with overdensities in the dust by a factor of three. This is the right range of metallicity and size of particles which is needed for the streaming instability to set in \citep{2005ApJ...620..459Y}. This instability can further enhance the clumping until gravity is strong enough to directly form planetesimals out of the cluster of particles.

\begin{figure}[t]
    \centering
    \includegraphics{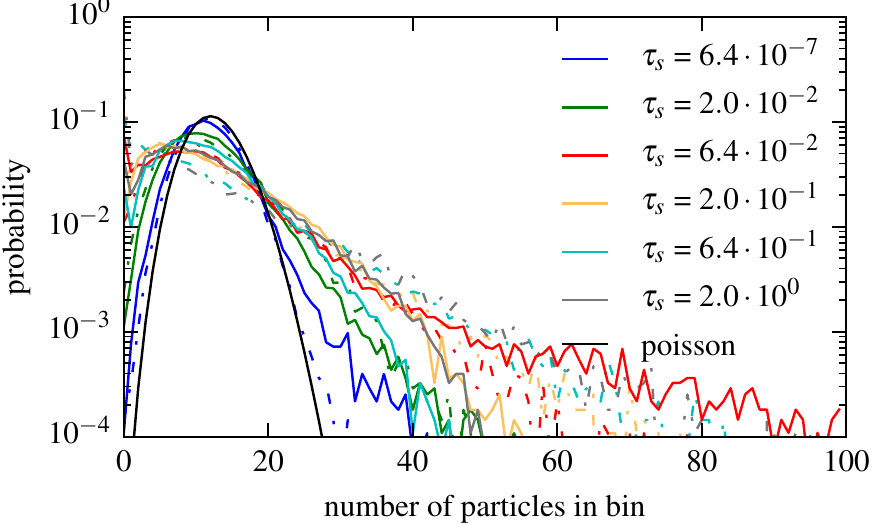}
    \caption{Probability to find a certain number of particles in a radial bin with $\Delta r = 0.05\AU$. The dashed-dotted lines correspond to the isothermal model.}
    \label{fig:irr_poisson}
\end{figure}

In Fig. \ref{fig:irr_poisson} we repeat the statistical analysis for the distribution of particles. We take into account all particles in the region from 15 to $40\AU$, in the timespan between 11000 and 13500 years over 50 snapshots. Note that the average number of particles per bin is the same as in the isothermal case in the earlier section, since we increased the size of the bins to compensate for the lower density of particles. Again we see a clear deviation from the initial Poisson distribution for the particles with stopping time around unity, even though it is weaker. Interestingly the effect is now most powerful for $\tau_s = 6.3 \cdot 10^{-2}$ (particles with \unit[10]{cm} radius), where the inward drift velocity for the particles and the inward drift of the vertical motion of the VSI mode is the same, thus the particles move with the bunching gas mode instead of through the mode.  Those are only bunched at around $30\AU$, and are diffused again after they have passed this region. This resonance does not exist in the isothermal case, because the wavelength is larger than in the radiative case. The isothermal case in general behaves very similar to the isothermal case in the earlier section. Both show the strongest bunching for particles with stopping time close to unity.

\begin{figure}[t]
    \centering
    \includegraphics{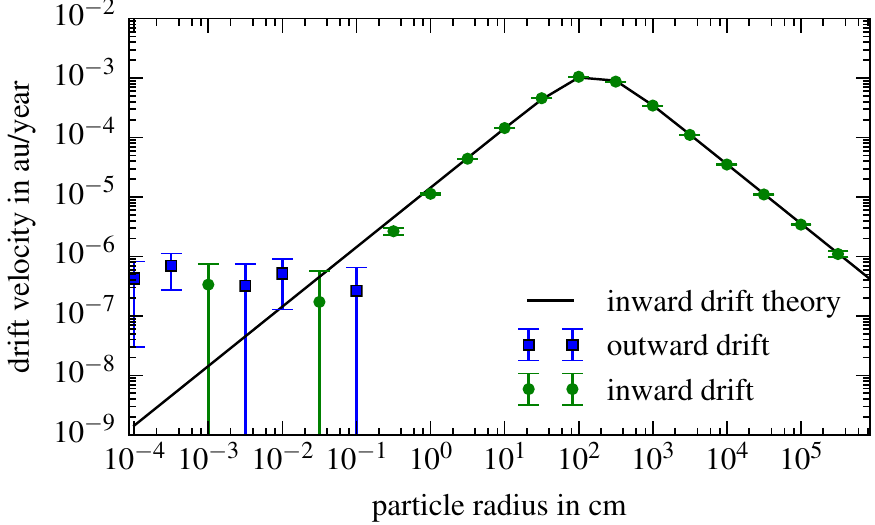}
\caption{The drift velocity of the dust particles depending on the radius of the particles. This is compared to expected transport for pressure supported discs. 
  Results are shown for the radiative simulation with irradiation, resolution $1024\times256\times64$ and viscosity $\nu=5 \cdot 10^{-7}$ at $r=20\AU$.
  Different colors are used for inward and outward drift. We estimate the error from the radial diffusion coefficent.}
    \label{fig:irr_mig}
\end{figure}

The radial drift shown in Fig. \ref{fig:irr_mig} measured at $20\pm 2\AU$ from 13500 years to 18500 years is similar to the isothermal case with the same viscosity. While the outward migration is no longer as clear as in the isothermal case with viscosity, there is still a trend to outward migration. That the effect is weaker can be explained by the weaker effect the viscosity has at 20\AU.

\begin{figure}[t]
    \centering
    \includegraphics{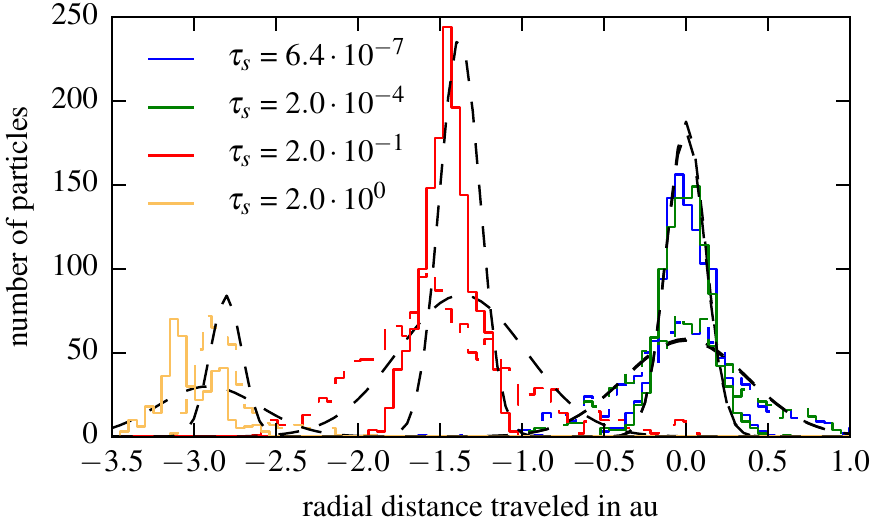}
\caption{Radial diffusion after 11000 years for 500 years for the radiative simulation with irradiation and viscosity $\nu = 5 \cdot 10^{-7}$. The black dashed lines are calculated from theory for the different stopping times and with the same $\tau_{eddy} = 1.0$ and $<u_{gas,radial}^2> = 5 \cdot 10^{-8} \cdot u_{Kepler,1au}^2$. The isothermal simulation (dashed lines) has a velocity dispersion of $<u_{gas,radial}^2> = 5 \cdot 10^{-7} \cdot u_{Kepler,1au}^2$. }
    \label{fig:irr_radDiff}
\end{figure}

More important for the radial motion of a single particle is the diffusion.
For the radial and vertical velocity dispersion in the region at $20 \pm5\AU$ we measure for the radiative case $<u_{r}^2> = 5 \cdot 10^{-8} \cdot u_{Kepler,1au}^2$ and $<u_z^2> = 5 \cdot 10^{-7} \cdot u_{Kepler,1au}^2$ and for the isothermal case we measure $<u_{r}^2> = 5 \cdot 10^{-7} \cdot u_{Kepler,1au}^2$ and $<u_z^2> = 2 \cdot 10^{-6} \cdot u_{Kepler,1au}^2$. 
Both values lead to a dimensionless eddy time of $\tau_{eddy} = 1.0$ even though the velocity dispersion differs by a factor of ten. The larger difference between prediction and simulation in Fig. \ref{fig:irr_radDiff} results from the error in the measurement of $q$, the exponent in the radial temperature distribution. Here, $T(r)$ is determined through the radiation transport and $q$ varies now with radius. For the plot we use an average value of $q = -1.1$.

In table \ref{tab:verdiff1} we can see that in this simulation the dust scale height is smaller than the gas scale height even for the smallest particles.  We averaged from 13500 years to 18500 years. In this simulation the radial and vertical calculated eddy lifetimes are again very similar, despite the turbulence not being isotropic.

\begin{table}
\begin{tabular}{cccc}
\hline \hline
radius & stopping time $\tau_s$& $h_{{p,z}}$ & $t_{{eddy,z}}$ \\
cm & at 20\AU & $h_{{gas}}$ & at 20\AU \\
\hline
$  1.0 \cdot 10^{ -4 }$ & $  6.3 \cdot 10^{ -7 }$ & $  9.4 \cdot 10^{ -1 }$ & - \\
$  3.2 \cdot 10^{ -4 }$ & $  2.0 \cdot 10^{ -6 }$ & $  9.4 \cdot 10^{ -1 }$ & - \\
$  1.0 \cdot 10^{ -3 }$ & $  6.3 \cdot 10^{ -6 }$ & $  9.5 \cdot 10^{ -1 }$ & - \\
$  3.2 \cdot 10^{ -3 }$ & $  2.0 \cdot 10^{ -5 }$ & $  8.0 \cdot 10^{ -1 }$ & $  1.3 \cdot 10^{ -2 }$ \\
$  1.0 \cdot 10^{ -2 }$ & $  6.3 \cdot 10^{ -5 }$ & $  8.7 \cdot 10^{ -1 }$ & $  7.3 \cdot 10^{ -2 }$ \\
$  3.2 \cdot 10^{ -2 }$ & $  2.0 \cdot 10^{ -4 }$ & $  8.7 \cdot 10^{ -1 }$ & $  2.4 \cdot 10^{ -1 }$ \\
$  1.0 \cdot 10^{ -1 }$ & $  6.3 \cdot 10^{ -4 }$ & $  6.6 \cdot 10^{ -1 }$ & $  1.8 \cdot 10^{ -1 }$ \\
$  3.2 \cdot 10^{ -1 }$ & $  2.0 \cdot 10^{ -3 }$ & $  5.4 \cdot 10^{ -1 }$ & $  3.1 \cdot 10^{ -1 }$ \\
$  1.0 \cdot 10^{ 0 }$ & $  6.3 \cdot 10^{ -3 }$ & $  4.2 \cdot 10^{ -1 }$ & $  5.0 \cdot 10^{ -1 }$ \\
$  3.2 \cdot 10^{ 0 }$ & $  2.0 \cdot 10^{ -2 }$ & $  3.4 \cdot 10^{ -1 }$ & $  9.9 \cdot 10^{ -1 }$ \\
$  1.0 \cdot 10^{ 1 }$ & $  6.3 \cdot 10^{ -2 }$ & $  2.3 \cdot 10^{ -1 }$ & $  1.3 \cdot 10^{ 0 }$ \\
$  3.2 \cdot 10^{ 1 }$ & $  2.0 \cdot 10^{ -1 }$ & $  8.9 \cdot 10^{ -2 }$ & $  5.9 \cdot 10^{ -1 }$ \\
$  1.0 \cdot 10^{ 2 }$ & $  6.3 \cdot 10^{ -1 }$ & $  3.1 \cdot 10^{ -2 }$ & $  2.3 \cdot 10^{ -1 }$ \\
$  3.2 \cdot 10^{ 2 }$ & $  2.0 \cdot 10^{ 0 }$ & $  1.5 \cdot 10^{ -2 }$ & $  1.6 \cdot 10^{ -1 }$ \\
$  1.0 \cdot 10^{ 3 }$ & $  6.3 \cdot 10^{ 0 }$ & $  4.1 \cdot 10^{ -3 }$ & $  3.9 \cdot 10^{ -2 }$ \\
\hline
\end{tabular}
\caption{Measured dust scale heights and inferred eddy lifetimes, from Eq.~(\ref{eq:hdust}), for the radiative model.
We fitted a Gaussian $f(z) = N_0 \exp{(-(z\pm\mu)^2/(2 r^2 h_{p}^2 })$ to the data of the vertical distribution to find $h_{p}$ and calculated $\tau_{eddy}$ for the irradiated simulation with viscosity of $5\cdot 10^{-7}$. }
\label{tab:verdiff1}
\end{table}

\begin{figure}[t]
    \centering
    \includegraphics{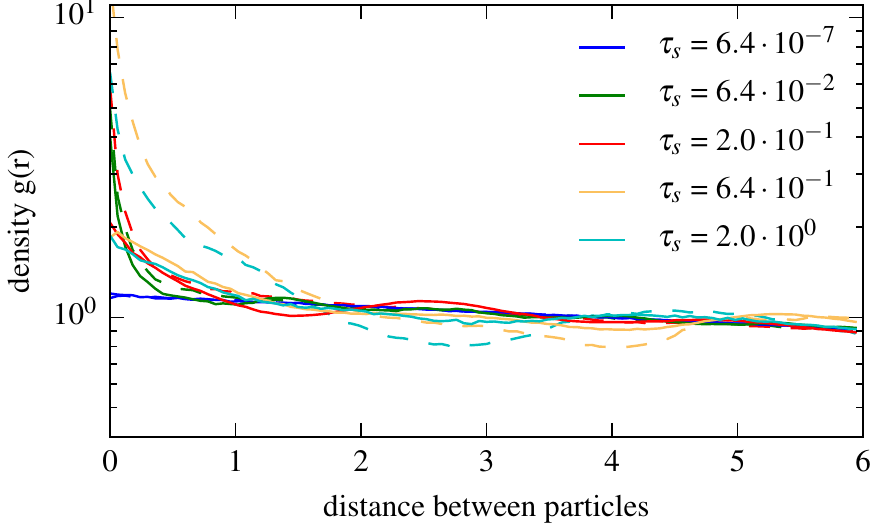}
    \caption{Pair correlation function for the irradiated simulation: enhancement of surface density in a shell with radius $r$ for different stopping times. Averaged over 5 snapshots  in the interval from 11000 to 13500 years. The dashed lines correspond to the isothermal model.}
    \label{fig:irr_corr}
\end{figure}

For the collision statistics we increased the cutoff distance within which we compare particle velocities to $0.2\AU$ to compensate the decreased density of particles. As the distribution of particles already indicated the clustering is indeed weaker. This is reflected additionally in Fig. \ref{fig:irr_corr}, where we can see that the correlation is slightly weaker. 
For this radiative case the effect appears to be strongest for a dimensionless stopping time $\tau_s \approx 0.1$ instead of 1 for the isothermal case. The correlation length is larger for the particles with stopping time around one, reflecting the larger wavelength of the VSI in the isothermal case. 

\begin{figure}[t]
    \centering
    \includegraphics{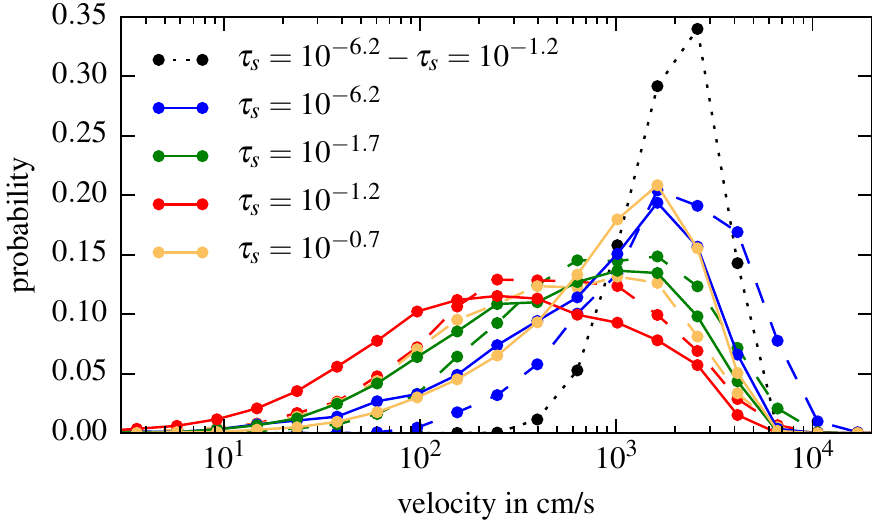}
    \caption{Histogram: Relative velocity between particles of the same size for different stopping times after 11000 years for the irradiated simulation. 
The dotted lines correspond to two different particle sizes as indicated in the legend. 
The dashed lines correspond to the isothermal model. 
Lines to guide the eye.}
    \label{fig:irr_colVel}
\end{figure}

The histogram of the relative velocities between particles as displayed in Fig. \ref{fig:irr_colVel} illustrates this situation. For $\tau_s \approx 0.1$ the particles have about an order of magnitude smaller relative velocities than for large and small values. We can also see that the larger velocity dispersion in the isothermal model leads to larger relative velocities. 

\section{Summary and conclusions}\label{conclude}
%
In the paper we analyzed the dynamics of particles embedded in hydrodynamic discs that show
fully developed turbulence as induced by the VSI.

In a first step we calculated isothermal disc models in full three dimensions and analyzed the properties
of the turbulence generated by the VSI. Our standard model consisted of an eighth of a full circle ($\phi_{max} = \pi/4$)
and showed in the fully developed turbulent state $\alpha$-values around $6 \cdot 10^{-4}$, which is of the same order of magnitude
or even slightly larger than the corresponding 2D models \citep{Stoll2014A&A...572A..77S}. The 3D models shows
variations in the azimuthal direction and these fluctuations follow a Kolmogorov-type spectrum. The mean radial velocity of the gas 
in a VSI turbulent disc turned out to be directed inward in the disc midplane and outward in the upper layers,
in agreement with global MHD simulations using zero net vertical magnetic flux
\citep{2011ApJ...735..122F}. This flow is opposite to viscous laminar discs \citep{1984SvA....28...50U,1992ApJ...397..600K} or MHD discs with non-zero vertical
magenetic field \citep{2014ApJ...784..121S}.
For 3D discs covering the full circle ($\phi_{max} = 2 \pi$) we found very similar results, which allowed us to treat particle
evolution in the reduced domain. 

In addition to the isothermal case we studied fully radiative models including heating from the
central star. To allow for regimes where the VSI instability can operate we extended to radial domain from
$8-80$\AU. The temperature structure in the disc displayed a central disc region with a nearly constant temperature in the vertical
direction and hotter surface layers produced by the stellar irradiation. The vertically varying opacity in the disc resulted in different
cooling times and the turbulence turned out to be slightly weaker in comparison to the purely isothermal situation. 
For the effective $\alpha$-parameter values of around $10^{-4}$ were reached in the active state that extended from about 10 to 60 \AU.

After having reached the equilibrium state we inserted particles of different sizes to study their motion in the disc, where
the drag force between gaseous disc and particles was treated in the Epstein regime. Overall we found for both, isothermal and radiative discs
comparable results. On average the particles drift inwards with the expected speed. For all disc models we found that the smallest 
particles show an outwardly directed radial drift. This comes about because the small particles are coupled more to the gas flow and are lifted
upward by the vertical motions of the VSI induced large scale flows. Since the average flow direction in the upper layers is positive
small dust particles that are elevated above the disk's midplane are dragged along and move outwards.
Particles below about 1~mm in size experience this fate. This outward drift might be beneficial in transporting
strongly heated solid material to larger radii as required to explain for example the presence of chondrules at larger radii in the Solar
System \citep{2002A&A...384.1107B}.
The upward drift of small particles in the disc by the VSI modes will also help to explain the observed presence of a population small particles
in the later stages of the disc evolution that were produced by a fragmentation process \citep{2005A&A...434..971D}.

Using the information of histograms, probability functions and pair correlation functions we analyzed the spatial re-distribution of particles
in the disc that were initially homogeneously distributed. We found that the particles are strongly 'bunched' together by the large scale motions of the
VSI turbulence. The bunching effect is strongest for particles with a stopping time of the order unity and the maximum overdensities
reached were about 5 times the average initial density of the particles. The relative velocity between particles of the same size is smallest
(about a few m/s) for those particles that show the strongest bunching. This combination of high density and low relative speed is highly beneficial
for the early formation process of planetary precursors. First, at these relative speeds collisions between two particles can lead to 
sticking collisions \citep{2008ARA&A..46...21B,2013MNRAS.435.2371M}. 
The higher relative velocities between particles of different sizes does not necessarily lead to fragmentation. The experiments of \citet{2009MNRAS.393.1584T} have shown that particles with different size can stick to each other even for collisions up to 50 m/s and possibly more.
Secondly, through the concentration of particles it is possible to
trigger streaming instabilities in the disc which can further increase the particle concentration and growth \citep{2005ApJ...620..459Y}.

The two dimensional distribution of particles in the disc shows axisymmetric ring-like concentration zones of the particles resembling
very roughly the features observed recently in the disc around HL~Tau \citep{Brogan2015ApJ...808L...3A}. 
Even though the strongest effect is seen here in our simulations for particles about one meter in size, 
it is possible that through collisions of nearly equal sized bodies much smaller particles that could generate the observed
emission can be produced and which follow a similar spatial distribution.
Obviously the observed spacing of the 'bright' rings in our simulations is smaller than those observed in HL~Tau but the
inclusion of variations in opacity or chemical abundances may create larger coherent structures.

\begin{acknowledgements}
Moritz Stoll received financial support from the Landesgraduiertenf\"orderung
of the state of Baden-W\"urttemberg and through the German Research Foundation (DFG) grant KL 650/16.
Simulations were performed on the bwGRiD cluster in T\"ubingen or the ForHLR cluster in Karlsruhe (both funded by the DFG and
the Ministry for Science, Research and Arts of the state Baden-W\"urttemberg), and the cluster
of the Forschergruppe FOR 759 “The Formation of Planets: The Critical First
Growth Phase” funded by the DFG.
\end{acknowledgements}

\bibliographystyle{aa}
\bibliography{library,refs,calibre}{}

\begin{appendix}
    \section{Particle Solver}
    \label{appsect:particles}
    To verify the correct implementation of our particle solver we repeat some of the tests of \citet{Zhu2014ApJ...785..122Z}.
    \subsection{Orbit test}
\begin{figure}[h]
        \centering
        \includegraphics{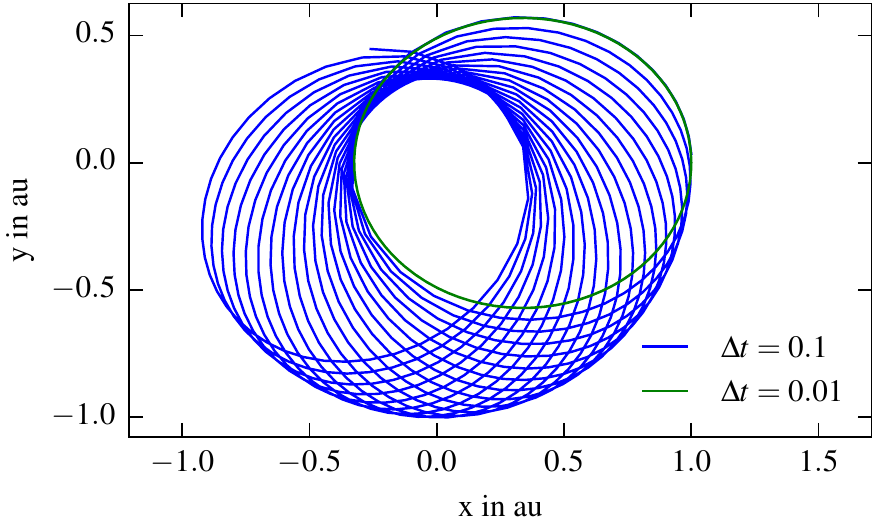}
        \caption{The orbital evolution of a test particle on an eccentric orbit.}
        \label{fig:orbit}
    \end{figure}

    We release the particle at $r = 1$, at the midplane with a velocity of $u_\phi = 0.7$ and integrate for 20 orbits. The presented timesteps are $\Delta t = 0.1 $ and $\Delta t =0.01$ for an orbital time of $2 \pi$. Even though the orbit precesses for the larger timestep, the geometric property is conserved. There is no visible precession in Fig. \ref{fig:orbit} for the timestep of $\Delta t = 0.01$, which we use in our simulations.

    \subsection{Settling test}
\begin{figure}[h]
        \centering
        \includegraphics{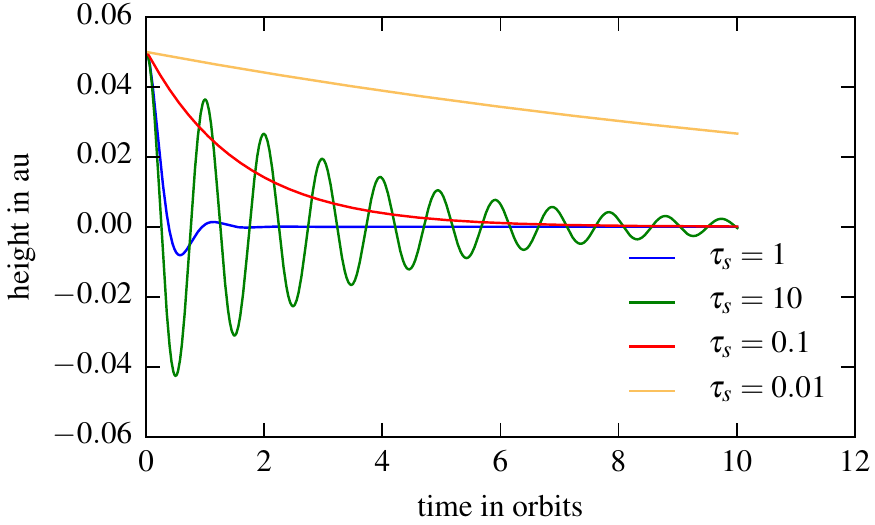}
        \caption{The settling of test particles with different stopping times.}
        \label{fig:settling}
    \end{figure}

    We release particles with different stopping times at one scale height from the midplane. For particles with $\tau_s < 1$ we can see in Fig. \ref{fig:settling} the exponential decay of the vertical position. Particles with $\tau_s > 1$ oscillate around the midplane and instead the amplitude decays exponentially.

    \subsection{Drift test}
\begin{figure}[h]
        \centering
        \includegraphics{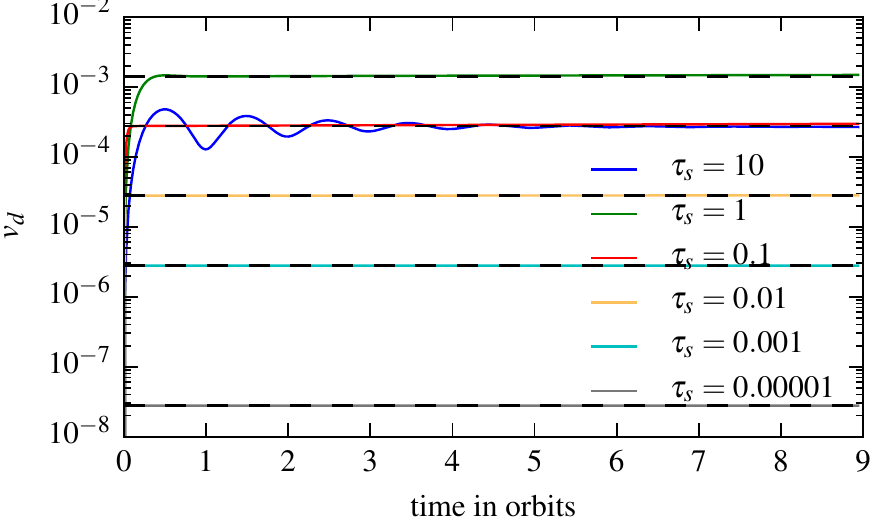}
        \caption{The drift velocity for particles with different stopping times.}
        \label{fig:drifttest}
    \end{figure}

    For the drift test we use the disc in hydrostatic equilibrium and release particles with different stopping times at $r = 5$au on Keplerian orbits in Fig. \ref{fig:drifttest}. We compare to the theoretical expected drift velocity of Eq. (\ref{eq:radialdrift}) (black line).  

\end{appendix}
\end{document}